\documentclass[aps,prb,reprint,superscriptaddress]{revtex4-2}

\pdfoutput=1
\usepackage{amssymb}
\usepackage{graphicx}
\usepackage{dcolumn}
\usepackage{bm}
\usepackage{amsmath}
\usepackage{amsfonts}
\usepackage{color}

\begin{document}
\title{Percolation in metal-insulator composites of randomly packed spherocylindrical nanoparticles}
\author{Shiva Pokhrel}
\thanks{These authors contributed equally to this work.}
\author{Brendon Waters}
\thanks{These authors contributed equally to this work.}
\affiliation{Department of Physics and Astronomy, Wayne State University, Detroit, Michigan 48201, USA}
\author{Solveig Felton}
\affiliation{Centre for Nanostructured Media, School of Mathematics and Physics, 
Queen's University Belfast, Belfast, BT7 1NN, UK}
\author{Zhi-Feng Huang}
\email{huang@wayne.edu}
\author{Boris Nadgorny}
\email{nadgorny@physics.wayne.edu}
\affiliation{Department of Physics and Astronomy, Wayne State University, Detroit, Michigan 48201, USA}
\date{\today}

\begin{abstract}
While classical percolation is well understood, percolation effects in 
randomly packed or jammed structures are much less explored. Here we investigate 
both experimentally and theoretically the electrical percolation in a binary 
composite system of disordered spherocylinders, to identify the 
relation between structural (percolation) and functional properties of
nanocomposites. Experimentally, we determine the percolation threshold 
$p_c$ and the conductivity critical exponent $t$ for composites of conducting 
(CrO$_2$) and insulating (Cr$_2$O$_3$) rodlike nanoparticles that are 
nominally geometrically identical, yielding $p_c=0.305 \pm 0.026$ 
and $t=2.52 \pm 0.03$ respectively. Simulations and 
modeling are implemented through a combination of the mechanical contraction 
method and a variant of random walk (de Gennes ant) approach, in which charge 
diffusion is correlated with the system conductivity via the Nernst-Einstein 
relation. The percolation threshold and critical exponents identified 
through finite size scaling are in good agreement with the 
experimental values. Curiously, the calculated percolation threshold 
for spherocylinders with an aspect ratio of 6.5, $p_c=0.312 \pm 0.002$, 
is very close (within numerical errors) to the one found previously in two 
other distinct systems of disordered jammed spheres and simple cubic lattice, 
an intriguing and surprising result.

\end{abstract}

\maketitle

\section{Introduction}

The concept of percolation enables a connection between the long-range
connectivity of randomly distributed objects within a network to global
properties of the system spanned by this network. The behavior of such systems
can be described by a standard model in which a random probabilistic process
shows a continuous phase transition from a finite size percolating cluster
below a critical value of the percolation threshold $p_{c}$ to an infinite
cluster above $p_{c}$ \cite{Stauffer}. Percolation is indispensable in
interpreting a wide variety of physical, chemical, mechanical, and biological
phenomena occurring in disordered systems, from the spread of diseases
\cite{epidemic}, thermal transport \cite{thermal}, electrical conduction in
composites \cite{Diffuse}, to metal-insulator \cite{electrical},
magnetic \cite{magnetic}, and spin quantum Hall \cite{spinHall}
phase transitions, and to pharmaceutical drug delivery \cite{Pharma1,Pharma2}.
Electrical conductivity in a percolating system can be modeled by
progressively adding larger numbers of identical conducting particles to an
insulating matrix until a geometrically connected conducting phase is
generated. The electrical conductivity $\sigma$ then scales as
$\sigma \propto (p-p_{c})^{t}$, where $p_{c}$ is the percolation threshold,
the critical value of the concentration or fraction $p$ of the
conducting particles, and $t$ is a critical exponent. The percolation 
threshold is normally dependent on the specific system configuration and 
the geometry of constituent particles. On the other hand, the critical 
exponent $t$ was expected to be universal, i.e., independent of details 
of system structures and components ($t = \mu \simeq 2$ in three dimensions 
\cite{Stauffer}), whereas more recent studies indicated the nonuniversality
of $t$, the values of which were found to range from $1.3$ to $4.0$ or even
higher in various composites \cite{BauhoferCST09,Vionnet-MenotPRB05,BalbergPRL17}.

\begin{figure}[ptb]
\centerline{\includegraphics[width=0.2\textwidth]{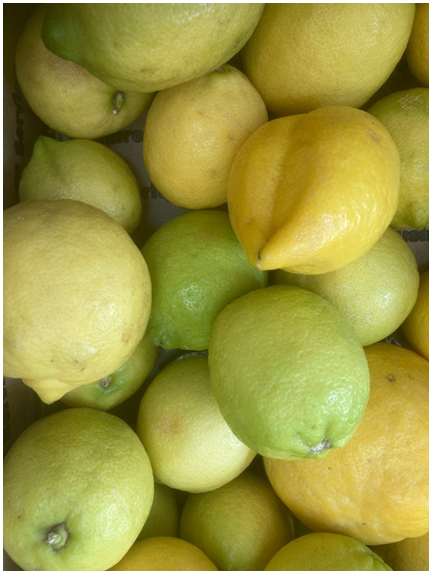}
\includegraphics[width=0.27\textwidth]{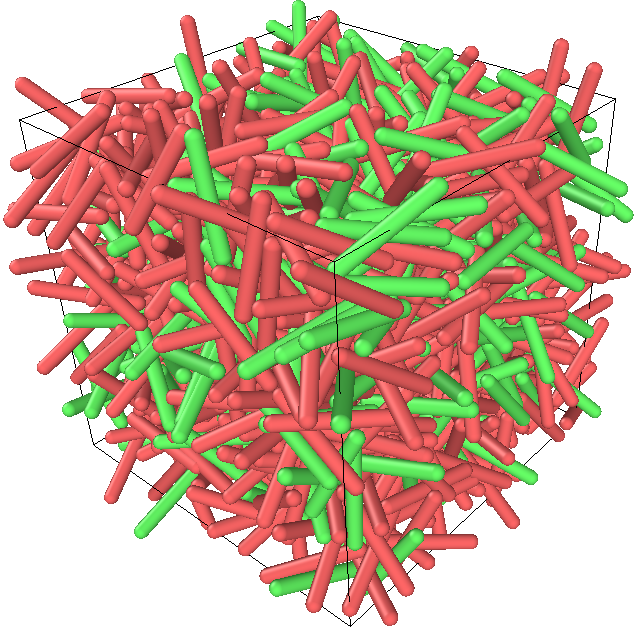}}
\centerline{\includegraphics[width=0.475\textwidth]{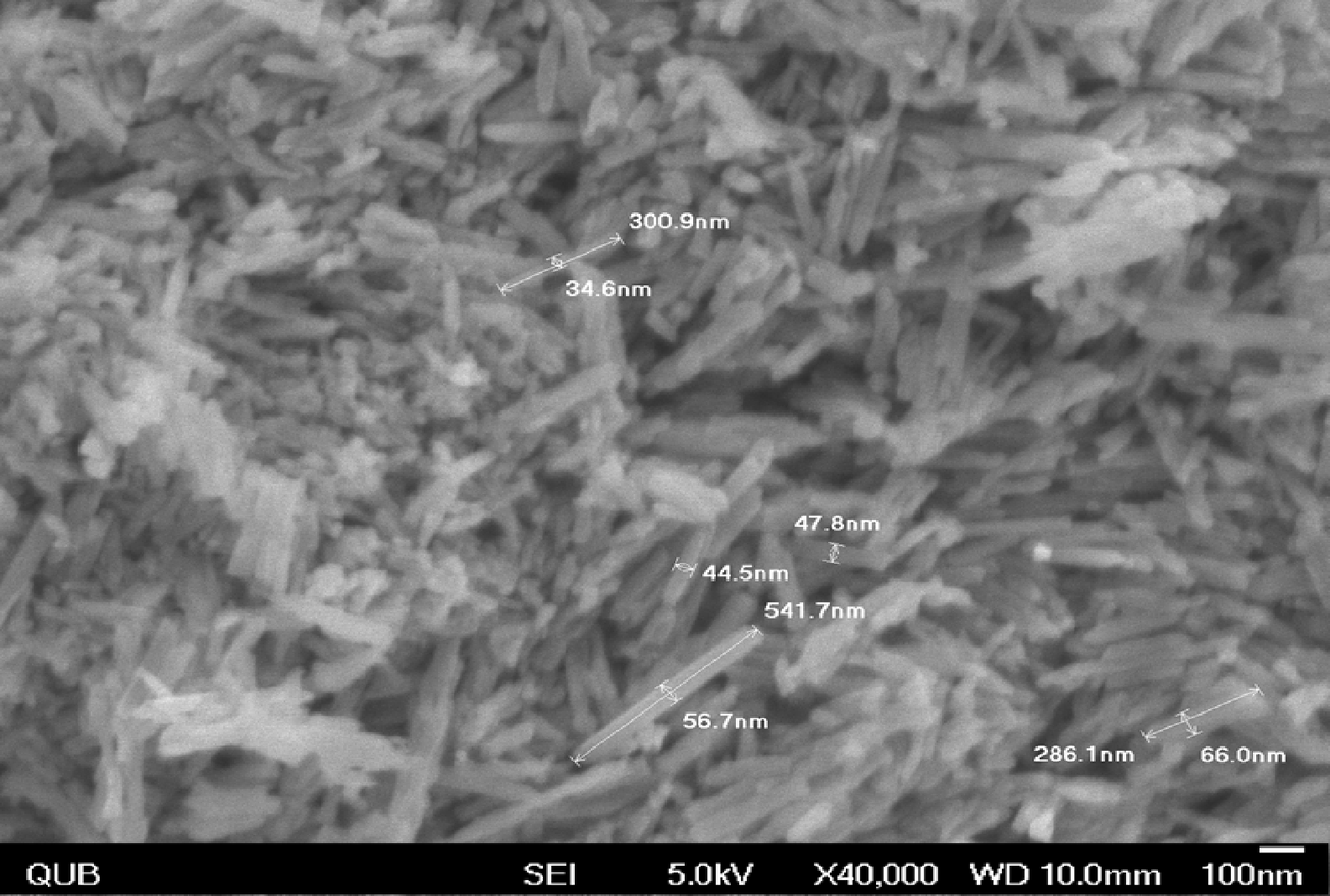}}
\caption{Sample systems of disordered packing, including closely packed 
homegrown lemons (top-left panel), a simulation snapshot of randomly packed
conducting (green, with $p=0.33$) and insulating (red) spherocylinders 
near the percolation threshold (top-right panel), and an SEM image of
the densely packed CrO$_{2}$/Cr$_{2}$O$_{3}$ experimental sample at approximately
the same fraction $p$ (bottom panel).
The average width and length of the CrO$_{2}$/Cr$_{2}$O$_{3}$ 
nanoparticles are measured to be approximately 40 nm and 300 nm, respectively.}
\label{Lemons}
\end{figure}

While the classical percolation picture described above has been well
established, a variant of this problem which addresses the percolation effects
of particles packed in disordered (random) or jammed structures is much less thoroughly
explored and understood \cite{Jamming} (see Fig.~\ref{Lemons} for some examples
of such disordered packings). Compared to lattice-like, ordered structures, 
randomly packed systems usually have different packing fractions,
which, in turn, would affect the critical behavior of the system and could
play a key role in defining the functionality of the sample.
A recent example is the occurrence of double percolation observed in a
disordered binary mixture \cite{double}, in which both types of
particles (CrO$_{2}$ and MgB$_{2}$) are conducting or superconducting but
their volume fraction vs conductivity relation shows an insulating region
in between two separate percolation thresholds. These two thresholds,
corresponding to the conductor-insulator and superconductor-insulator transitions
respectively, arose from the suppressed interface conduction between a half-metal 
(CrO$_{2}$) and a superconductor (MgB$_{2}$) \cite{dJ-B,Andreev} and the large 
geometric disparity between particles in this rod-sphere system of binary species
\cite{double}. This effect underscores the fundamental and practical importance
of the percolation threshold and the relationship between the thresholds,
geometric contrast of constituent particles, and the transport properties of 
the system for various particle networks.

While percolation thresholds and the critical behavior in many ordered three-dimensional
(3D) lattices of fixed coordination numbers have been investigated in detail
\cite{Stauffer,Diffuse,Thr1,Thr2,Thr3,Thr4,Thr5}, much less is known about systems
of disordered packing which form networks of interparticle contacts with variable
coordination numbers \cite{Z-T}. In these cases where the constituent particles
could be of various types and geometries, such as particle shape and size, and the
average number of nearest neighbors may be close to that in a specific ordered lattice, 
they are lacking long-range order and are randomly or quasi-randomly
distributed with a broad range of neighboring particle contact numbers.
To develop the methodology for identifying the corresponding percolation properties
(such as percolation thresholds and critical exponents) and their correlation to the
system functionalities, we will first limit ourselves to a more manageable problem
related to dense random distribution of particles of the same size and shape,
specifically the disordered mixture of two-component spherocylinders that are 
geometrically identical but functionally distinct, given that the understanding
of this type of system is still lacking.

In this paper, we report both experimental and computational results for a
randomly packed system comprising a binary mixture of nominally identical
CrO$_{2}$/Cr$_{2}$O$_{3}$ spherocylinders (capsules); see Fig.~\ref{Lemons}
for a simulation snapshot (top-right panel) and a scanning electron microscopy
(SEM) image of the system studied. Specifically, we study a dense disordered 
network of conducting and insulating spherocylinders of approximately the same 
size and aspect ratio. To examine the electrical percolation in this system, 
we first convert the conducting CrO$_{2}$ rodlike nanoparticles 
(spherocylinders with an average aspect ratio of $6.5$) into 
nominally geometrically identical insulating Cr$_{2}$O$_{3}$ nanoparticles, and
then prepare a series of samples of CrO$_{2}$/Cr$_{2}$O$_{3}$ mixture with varying
conducting vs insulating volume fractions for electrical transport measurements.
The experimental results are compared with large-scale 3D computer simulations
and calculations on the corresponding binary disordered composite of
non-overlapping hard spherocylinders, which are conducted through a combination
of mechanical contraction and Monte Carlo methods and a random walk approach
based on the de Gennes ant and the Nernst-Einstein relation.

Good agreement is obtained between our results from experiments and computation,
for various percolation and electrical transport properties of this binary network 
of disordered packing. These include the percolation threshold 
$p_c=0.305 \pm 0.026$ (experiment) and $p_c=0.312 \pm 0.002$ (simulation),
and the scaling behavior near the threshold and the corresponding critical
exponent of the electrical conductivity, with $t = 2.52 \pm 0.03$ and a
lower bound $\mu_l=1.26$ identified through $t \leq 2\mu_l$ \cite{BalbergPRL17}
(experiment) and $\mu = 1.62 \pm 0.04$ (simulation). The small discrepancy 
between the experimental and theoretical values can be partially attributed to 
the intrinsic polydispersity of the nanoparticles and some degree of local ordering
of the composite used in experiments (see Fig.~\ref{Lemons}).

An intriguing finding is the calculated value of $p_c=0.312 \pm 0.002$ in this binary 
system of dense randomly-packed (but not strictly jammed)
spherocylinders with an aspect ratio of $6.5$. 
This value is very close (within computational errors) 
to the site percolation threshold of $p_{c}=0.3116(3)$ computed recently
for disordered jammed spheres \cite{Z-T} and $p_{c}=0.311608$ obtained earlier for 
simple cubic lattice \cite{Thr1,Thr3}. These are three distinct systems, with different
packing fractions, degree of ordering, particle geometry, and different distribution
of particle coordination numbers (although with similar values of average coordination
numbers). While such an agreement may be accidental for any two systems, it is harder 
to assume that the same coincidence would occur for the three different systems, where 
the constraints of the ordered lattice and particle shape are being gradually removed. 
Hence, one may argue that this result manifests the existence of universality for the 
percolation threshold in a particular class of systems. Apart from pointing to the 
closeness of the average coordination numbers in all three systems, a plausible 
explanation of this result is still lacking; it remains an important open question.

\section{Experiments}

\subsection{Sample preparation and structural characterization}

Here experimental measurements are conducted to study the electrical transport
property of CrO$_{2}$/Cr$_{2}$O$_{3}$ half-metal/insulator nanocomposites at
different fractions of CrO$_{2}$. The CrO$_{2}$ and Cr$_{2}$O$_{3}$ nanoparticles 
are needle-shaped spherocylinders (see Fig. 1) of similar size,
with the average length of approximately 300~nm and diameter of about 40~nm,
corresponding to an average aspect ratio of $6.5$. 
Composite CrO$_{2}$/Cr$_{2}$O$_{3}$ samples were prepared by mixing metallic 
ferromagnetic oxide CrO$_{2}$ nanoparticles with insulating Cr$_{2}$O$_{3}$
nanoparticles. The latter were formed by annealing commercially available
CrO$_{2}$ (DuPont) nanoparticles at $550^{\circ}{\text C}$ for 1 hour, using the
procedure described in Ref.~\cite{Annealing}.

\begin{figure}
\begin{center}
\includegraphics[width=0.49\textwidth]{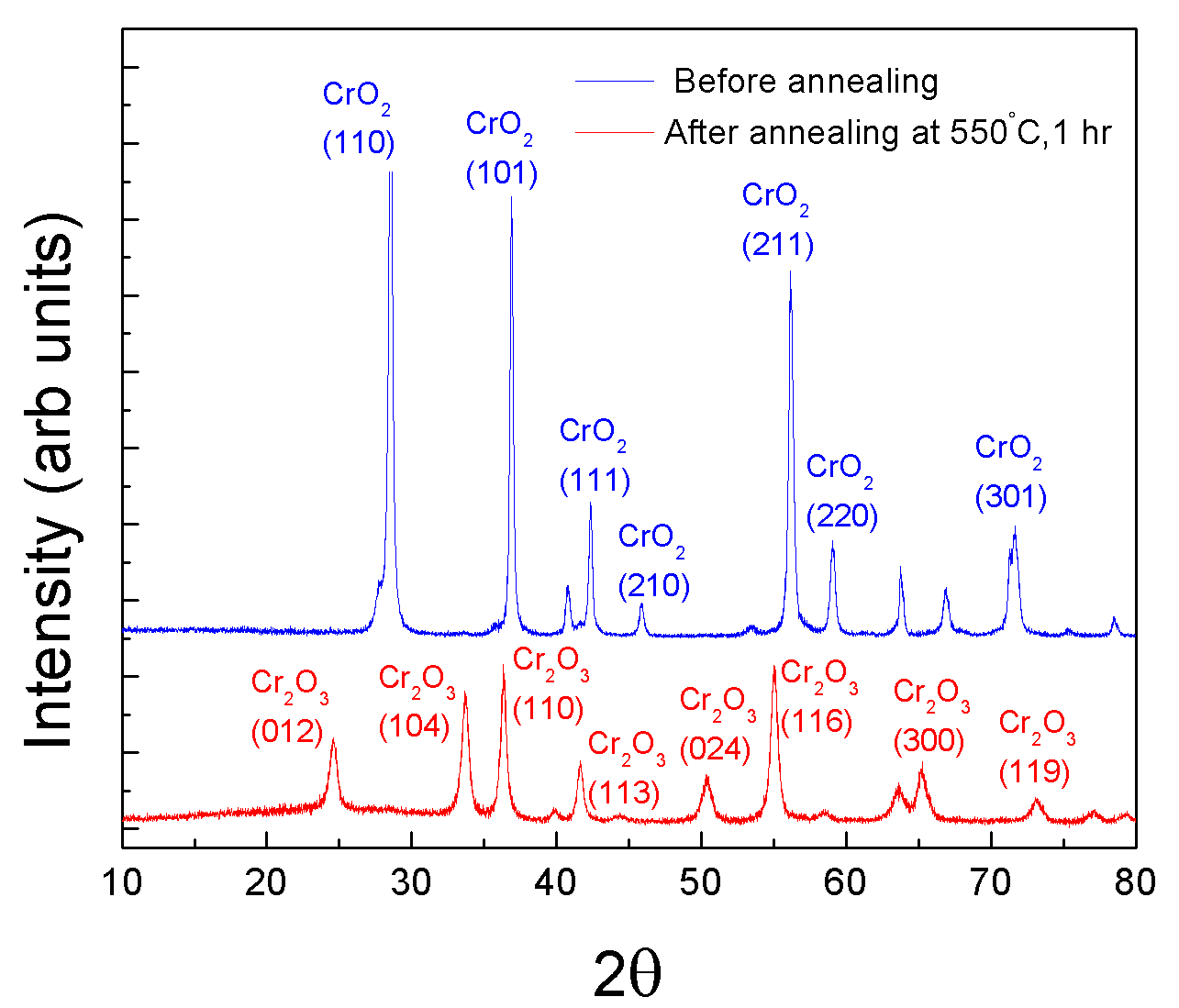}
\caption{The XRD intensity vs $2\theta$ plots, with the peaks corresponding to
the original CrO$_{2}$ nanoparticles before annealing and Cr$_{2}$O$_{3}$
nanoparticles after annealing.}
\label{XRAYS}%
\end{center}
\end{figure}

The resulting nanoparticles were then analyzed by x-ray diffraction (XRD)
spectroscopy to confirm the complete conversion of half-metallic CrO$_{2}$
into insulating Cr$_{2}$O$_{3}$ nanoparticles. XRD spectra over a wide range
of diffraction angles 2$\theta$ (varying from $10^\circ$ to $80^\circ$) was
collected with the help of an x-ray powder diffractometer (\textit{Rigaku Miniflex}),
using Cu $K\alpha$ radiation of wavelength $\lambda = 1.5418$~\AA.
The x-ray generator was operated at 40~kV with an anode current of
20~mA from stabilized power supplies. The scan was performed at a
scanning rate of $1^\circ$ per minute over the range of angles $10^\circ <
2\theta < 80^\circ$. Both annealed and unannealed samples of CrO$_{2}$ were
analyzed by XRD, with results given in Fig.~\ref{XRAYS}. Before annealing,
peaks appear at 2$\theta$ angles of approximately $28^\circ$, $36^\circ$,
$45^\circ$, ... These peaks correspond to the reflected intensity from the
(110), (101), (211), ... planes of the tetragonal structure of CrO$_{2}$.
Based on the Scherrer formula, the typical crystalline size $L_s$ of the CrO$_{2}$
nanoparticle is dependent on the wavelength $\lambda$, the line broadening
$B(2\theta)$ which is the full width at half-maximum, and the Scherrer constant
$K$ (at the order of 1): $L_s \simeq {K\lambda}/[B(2\theta)\cos\theta]
\sim 30$~nm, which is close to the diameter of the spherocylinder. 
This estimate is consistent with the result of the high resolution STEM
imaging (not shown), indicating good crystallinity of individual nanoparticles. 
Upon annealing, a clear signature of Cr$_{2}$O$_{3}$ peaks has been observed at
(012), (104), (024), ..., as shown in Fig.~\ref{XRAYS}, confirming the
presence of Cr$_{2}$O$_{3}$. This is accompanied by noticeably higher peak
broadening, indicating a reduced crystalline size of Cr$_{2}$O$_{3}$ nanoparticles.
Approximately 5~mm diameter pellets with 0.5~mm thickness were formed from
the mixture of conducting CrO$_{2}$ and insulating Cr$_{2}$O$_{3}$ particles,
using a cold-press die with a uniaxial pressure of 10~GPa.
An SEM micrograph of a typical sample is shown in Fig.~\ref{Lemons}.

\subsection{Electrical characterization}
\label{sec:experiment_results}

Four-point electrical transport measurements were performed in a Quantum
Design Physical Property Measurement System (PPMS), which allows a wide range of
resistance measurements at variable temperatures. Gold wires were attached to
the pellet using silver paste, and the sample was then mounted on a PPMS puck
using GE varnish. The variation of resistivity as a function of fraction
$p$ of CrO$_{2}$ particles in a CrO$_{2}$/Cr$_{2}$O$_{3}$ composite
was measured using PPMS in the temperature range 200~K $\leq T \leq $ 300~K.
At each nominal composition at least two different samples were
prepared. The resistivity was measured using the four-probe technique, 
which eliminates parasitic contributions from electrical contacts \cite{Fourprobe}. 
Experimental errors originated primarily from the geometric factors, such as 
the finite contact size and the contact placement uncertainty, as well as 
variations in the sample thickness and anisotropy. 
In addition to the classical percolation threshold reported here,
we have observed tunneling percolation thresholds at lower fractions $p$,
manifested by tunneling staircases similar to those described in 
Refs.~\cite{Stairs_Bal,Stairs}. The detailed results of these measurements 
will be presented elsewhere.

\begin{figure*}
\begin{center}
\includegraphics[width=0.325\textwidth]{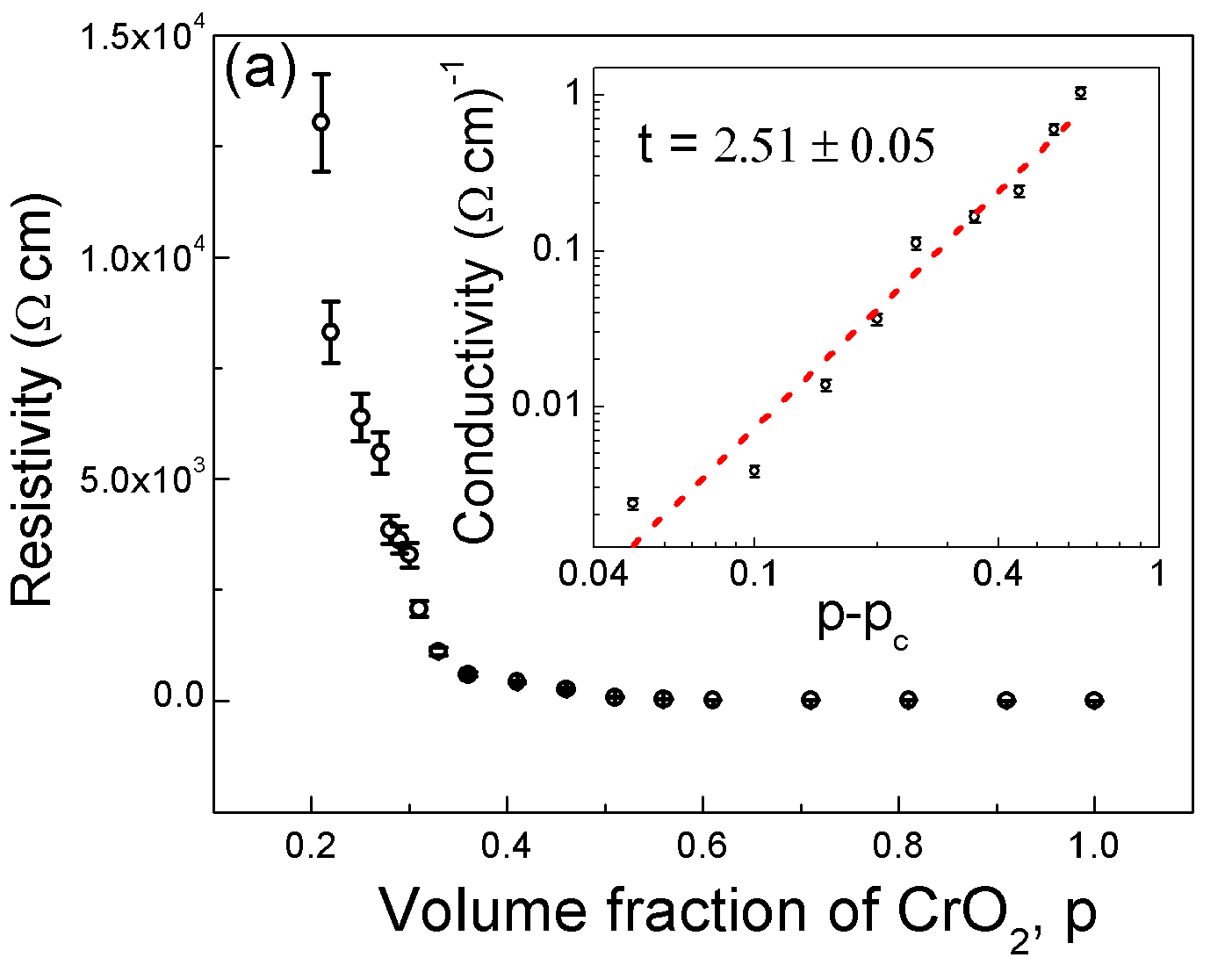}
\includegraphics[width=0.325\textwidth]{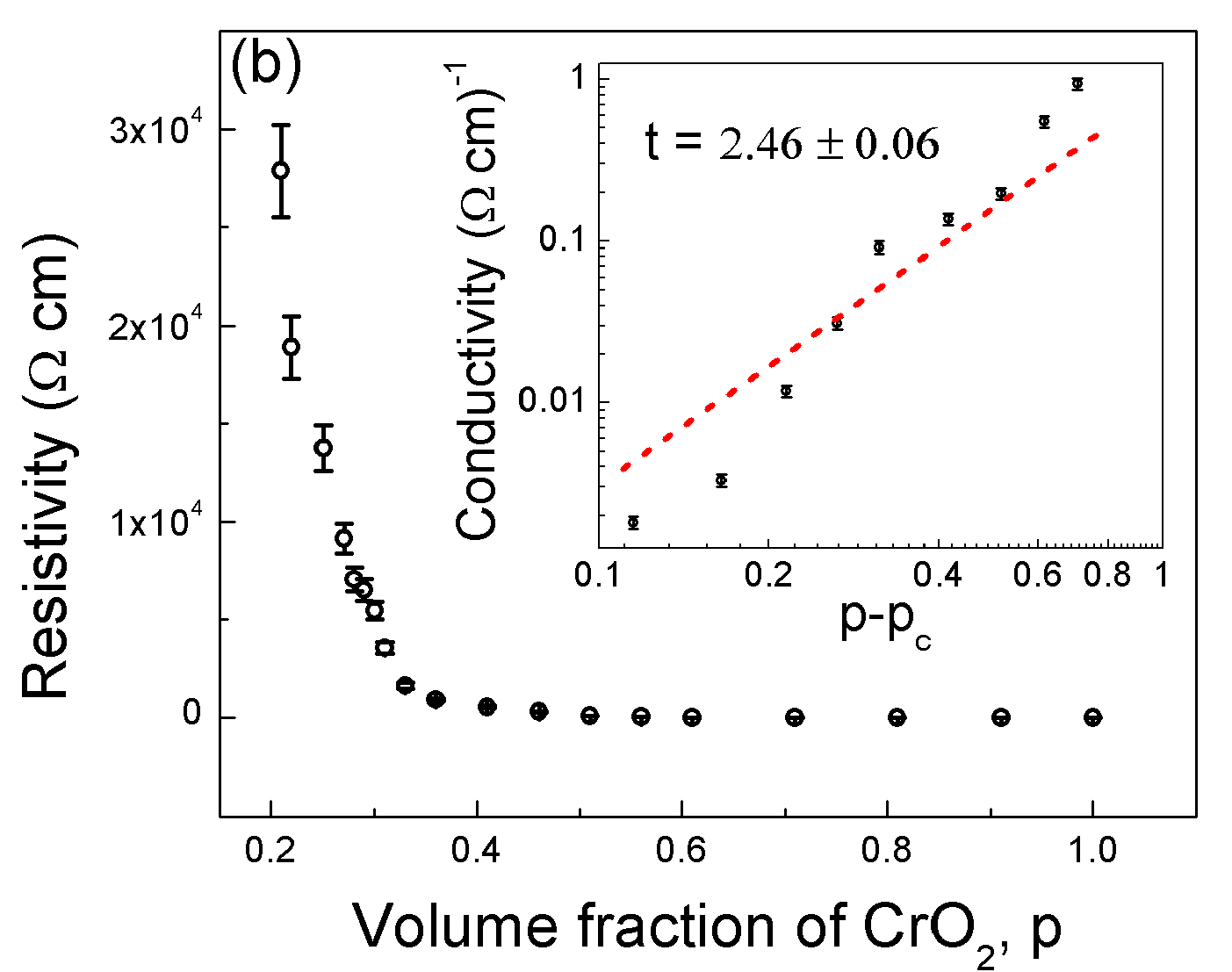}
\includegraphics[width=0.325\textwidth]{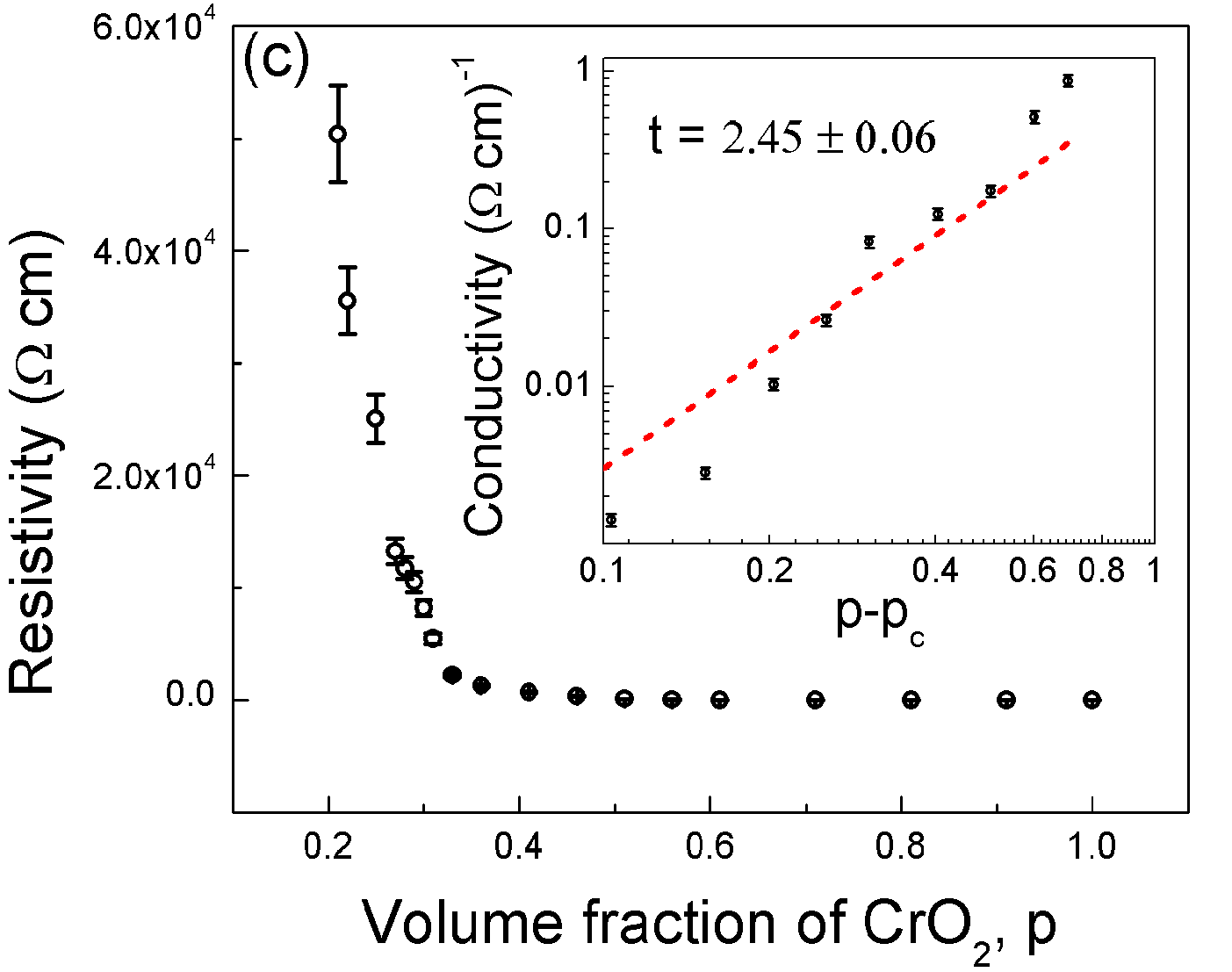}\\
\includegraphics[width=0.325\textwidth]{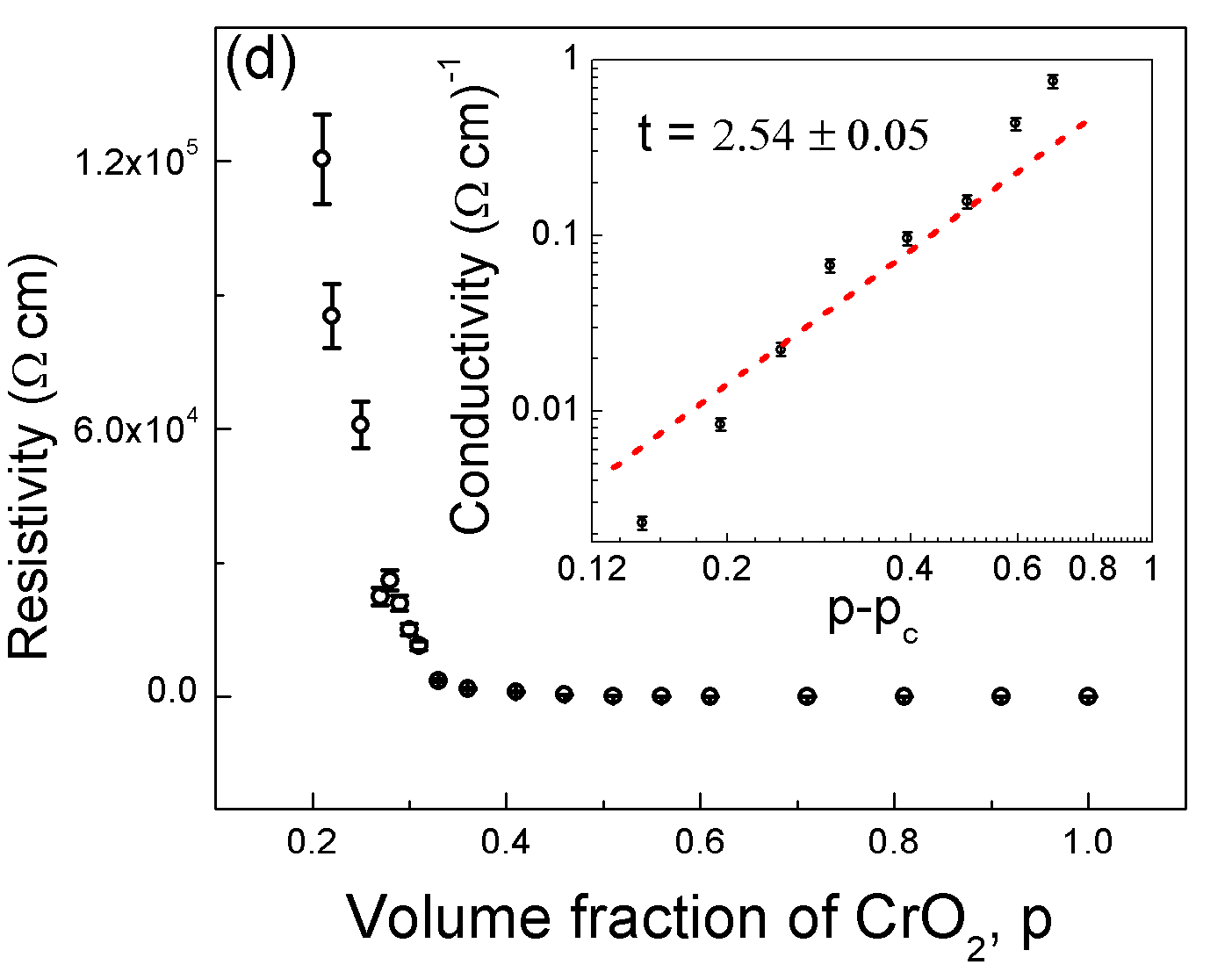}
\includegraphics[width=0.325\textwidth]{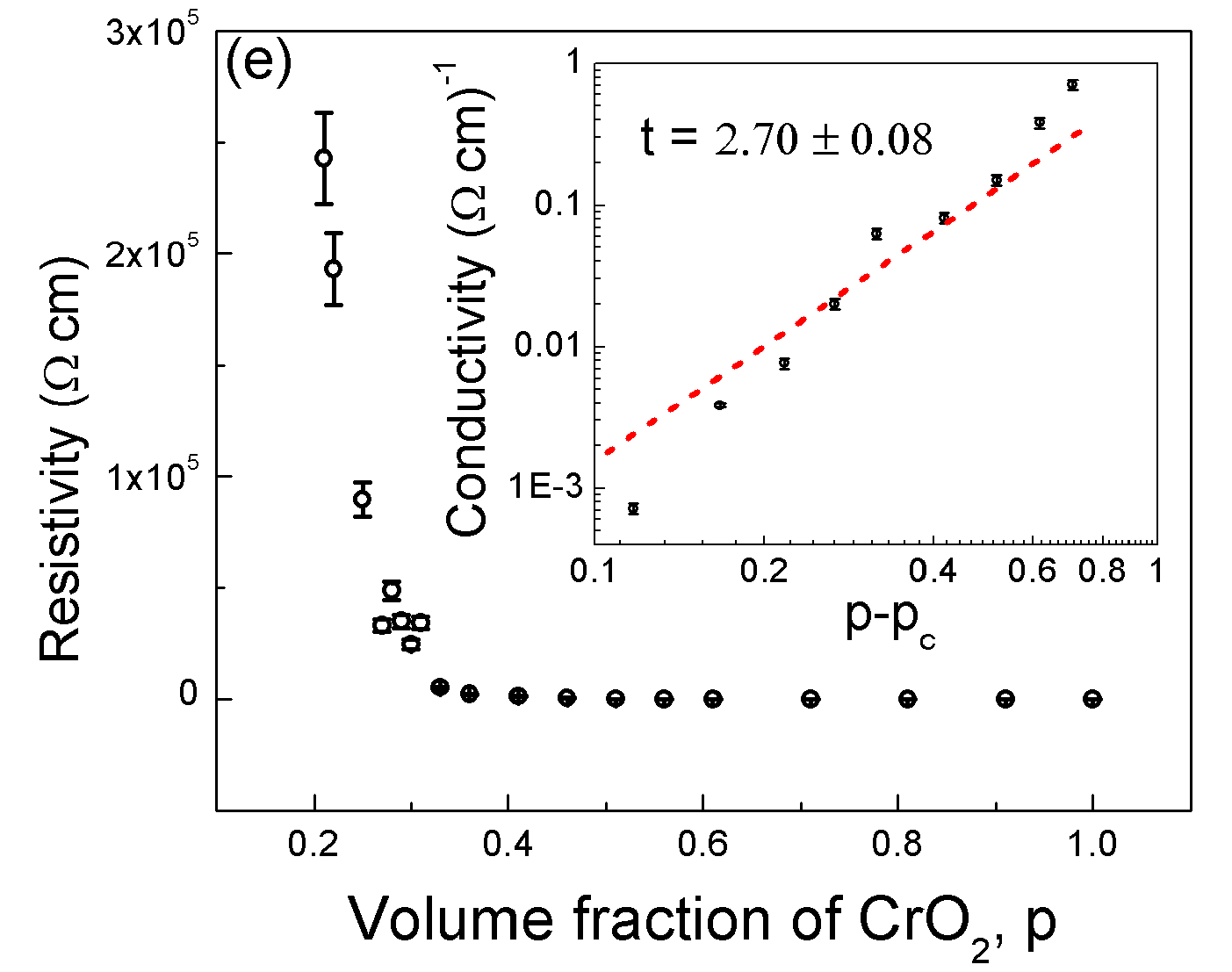}
\caption{Sample resistivity as a function of fraction $p$ of CrO$_{2}$ 
at five different temperatures (a) $T=300$~K, (b) $T= 270$~K, 
(c) $T=250$~K, (d) $T=220$~K, and (e) $T=200$~K. Insets: Sample conductivity 
as a function of $p-p_{c}$ in log-log scale to identify the critical exponent $t$
(where the error of $p_c$ determined from a separate nonlinear
fitting procedure has been incorporated).}
\label{R-p}
\end{center}
\end{figure*}

\begin{table}
\caption{Values of percolation threshold $p_c$ and conductivity
critical exponent $t$ obtained from experimental measurements at different 
temperatures $T$.}
\begin{ruledtabular}
\begin{tabular}{c c c}
$T$ & $p_c$ & $t$ \\ \hline
300 K & $0.360 \pm 0.024$ & $2.51 \pm 0.05$ \\
270 K & $0.295 \pm 0.057$ & $2.46 \pm 0.06$ \\
250 K & $0.307 \pm 0.062$ & $2.45 \pm 0.06$ \\
220 K & $0.315 \pm 0.040$ & $2.54 \pm 0.05$ \\
200 K & $0.293 \pm 0.062$ & $2.70 \pm 0.08$ \\
\end{tabular}
\end{ruledtabular}
\label{table_I}
\end{table}

Electrical characterization of nanocomposites for varying fractions of the 
conducting particles (CrO$_{2}$) was performed at five 
different temperatures, 300~K, 270~K, 250~K, 220~K, and 200~K, with the results 
shown in Fig.~\ref{R-p} and summarized in Table~\ref{table_I}. The values of 
the percolation threshold $p_c$ were determined by optimizing nonlinear 
regression for $\log \sigma = A + t \log (p - p_c)$ (plots not shown) from 
these five independent measurements of the sample conductivity $\sigma$, 
as listed in Table~\ref{table_I}.
Note that the conductivity of metallic
CrO$_{2}$ in this temperature range is only weakly temperature dependent, while
Cr$_{2}$O$_{3}$ components are at least an order of magnitude more resistive 
at $T = 250$~K as compared to room temperature (see Fig.~\ref{R-p}). 
Thus, the conductivity at $T = 300$~K is affected by the 
leakage current through Cr$_{2}$O$_{3}$ nanoparticles and hence was excluded 
from the determination of the $p_c$ and $t$ values of the conducting-insulating system. 
Averaging over the four lower temperature values of $p_c$, we identify the percolation 
threshold as $p_c=0.305 \pm 0.026$ for this nanocomposite system (using the proper 
weighted mean with the uncertainty evaluated via the standard error propagation). 
To perform the consistency check of these results, we used two supplementary 
methods for the threshold determination. In the first, the steepest descent 
(gradient) approximation \cite{ChenCPL03}, the threshold was associated with 
the extremum of the derivatives of $\log \sigma$ plotted as a function of $p$. 
In the second method, we used temperature-dependent transport measurements to 
identify inflection points in the plot of the ratios of sample resistances at a 
given temperature to the room-temperature values, as described in Ref.~\cite{Stairs}. 
The values of $p_c$ obtained from all these three methods coincide within the 
experimental accuracy.

Once the percolation threshold at a given temperature 
was identified, we used its value (with the corresponding error correction) in 
standard log-log plots to determine the conductivity critical exponent $t$. 
The results are shown in the insets of Fig.~\ref{R-p} (see also Table~\ref{table_I}), 
demonstrating the scaling behavior of conductivity above the percolation threshold 
$p_c$. The average of four lower temperature results yields $t= 2.52 \pm 0.03$.
This value of conductivity exponent for binary disordered spherocylinders is 
comparable to some previous results of 3D conductive networks, such as $t=2.16$ 
for the conductor-insulator transition in CrO$_2$/MgB$_2$ (spherocylinder/sphere)
double-percolating composites \cite{double} and those found in various disordered 
composites \cite{BauhoferCST09,Vionnet-MenotPRB05}.

\section{Simulations}

\subsection{Methods}

\subsubsection{The mechanical contraction method for random packing}

To model the structure of a given nanocomposite, it is essential to produce dense
random packings of its constituent hard particles in simulations. Conventional
Molecular Dynamics simulations are too computationally expensive to be practical
when the precise microstates of the system are not essential \cite{MD1997}.
Standard Monte Carlo methods inherently sample the equilibrium distribution of
particle states \cite{MC2006}, while making rapid compression into a disordered,
out-of-equilibrium state is not compatible with the premise of the model.
Additionally, this results in poor performance since in a dense
state an unacceptably high fraction of trial moves would fail the Metropolis
criterion \cite{Metro} for hard particles; that is, a vast majority of potential 
translations or rotations of particles produce states where particles overlap with
their neighbors. These states are invalid and thus must be rejected.

Because of these limitations, here we use an alternative algorithm designed
explicitly to produce random packings, the Mechanical Contraction Method
(MCM), as developed in Ref.~\cite{MCM}. The premise of MCM is to take a system
of spherocylinder particles in an initial low density random state, and bring
them together directly, moving each particle only when it would collide with
another and only just enough to avoid overlaps. By this approach, the initial
entropy of the low density phase is carried through the compression until the
entire system can no longer be transformed to a higher density state \cite{OtherMCM}.
Details of the method are given in Ref.~\cite{MCM}, with the implementation 
steps summarized below.

First, the particles' positions and orientations are randomized at low density
(with packing fraction $<0.01$) by a number of hard particle Monte Carlo moves.
At such low densities, virtually all trial moves succeed; thus this step
can be performed quickly, even for large systems. This provides the initial
state for the MCM. Using the MCM algorithm, the whole system is then scaled
down by a small volume factor $\Delta V$, and the position of every particle
is scaled down accordingly, bringing them uniformly together. The algorithm
searches over all particles in the system in arbitrary order, to identify
neighboring particles that might be overlapping. This is straightforward for
spherocylinders which can be characterized as a set of points within radius $R$
of a line segment of length $l$. If the shortest distance $k$ between the two
particles' axes of symmetry (within $l/2$ from the particle centroid) is within
$2R$, they must be overlapping. The amount of overlap is defined as $\delta=2R-k$.

The strategy of the MCM algorithm is to move a particle $i$ in a direction
that reduces the total of its overlaps with neighboring contacted particles
$j=1,2,...,C$ the most quickly. This direction is determined by maximizing
the effective speed, which is a weighted combination of translational and
rotational velocities with respect to all of its $C$ contacted particles
\cite{MCM}. Along this direction the particle $i$ is moved sufficiently far
to reduce the smallest overlap $\delta_{j}$ by a tiny amount more than
$\delta_{j}/2$, so that when particle $j$ is moved in the opposite direction
by the same amount the pair will barely break contact, typically separated
by $1.0001$ times the needed distance. This will minimize the probability of
producing new overlaps when moving the particles. This process is iterated 
until all overlaps are removed, whereupon the simulation
box is further reduced. The procedure is repeated until the system cannot
be further compressed with all the generated particle overlaps being removed
(within a large enough cutoff number of trials).

We have implemented the MCM algorithm as a custom module in HOOMD-Blue,
an open-source general purpose particle simulation engine
\cite{HOOMD1,HOOMD2,HOOMD_HPMC} that was used for generating the related
Monte Carlo moves in this algorithm (with excluded-volume interparticle
interaction). The custom module developed is available at
Ref.~\cite{MCM_github}. The acquired data were visualized using Ovito 
open-source particle visualization software \cite{ovito} (see e.g., 
Fig.~\ref{Lemons}).

\subsubsection{The random walk method for conductivity calculation}

Determining the conductivity of large networks of irregularly connected
particles presents a considerable challenge. To directly analyze the circuit
formed by conducting particles and use Kirchhoff's laws requires the 
construction and solution of a large set of linear equations, 
a procedure that is time consuming. Moreover,
the bulk conductivity of the system is likely insensitive to many small
changes in network structure. A more efficient approach can be obtained
by tracing a randomly-moving test charge through the system and calculating
its diffusion property, in a method analogous to the so-called de Gennes
``ant in a labyrinth'' \cite{deGennes} and ``termite'' \cite{deGennes_termite}, 
in concert with using the Nernst-Einstein relation. 

This random-walk model was initially proposed by de Gennes to explore the
percolation transition in cross-linked network systems \cite{deGennes},
where he imagined a microscopic ``ant'' lost in a labyrinth of nodes, some
of which are connected by bonds or chains. If these chains are sufficiently
cross-linked to create a percolating network, there should be a finite
probability that from any arbitrary starting point the ant could walk
along the network to cover infinite distance. Later theoretical work expanded
upon this simple percolation test and used this random-walk approach to
measure properties related to the diffusion processes, connectivity, and
transport in disordered systems or random networks \cite{Stauffer,HavlinAdvPhys87}.
These include the use of the de Gennes ant to calculate the diffusion
constant and hence the conductivity of the random resistor network
through the Nernst-Einstein relation, the extension to the de Gennes
termite \cite{deGennes_termite} to study the random superconducting
network consisting of normal conductors and superconductors, and further
to more complex cases of composites involving two \cite{Diffuse2} or three
\cite{Diffuse3,Diffuse4} types of bonds with different conductances.

Here we use this random walk approach to obtain the conductivity of the
disordered system of conducting-insulating spherocylinder nanoparticles, 
by imagining our ant as a test charge diffusing through the
particle network, subject to the local conductivity of each particle.
In practice, this is done by initializing the random walker on an arbitrarily
chosen particle in the system consisting of all the clusters
(i.e., the general ensemble \cite{HavlinAdvPhys87}), following 
the procedure for a ``blind ant'' \cite{MajidPRB84}. 
At each step, the walker (ant) located at particle $i$ chooses at random 
one of the neighboring particles $j$ which it is in contact with. 
The ant moves to that particle if it is conductive;
otherwise the location of the ant remains unchanged. In either case the amount
of time $t$ taken by the ant is increased by one unit at every step.
This corresponds to the assumption that in our system of 
CrO$_2$/Cr$_2$O$_3$ composites, only two types of particle-particle conductance 
are taken into account, with the interparticle hopping rates for a random 
walker being $1$ and $0$ respectively. The first case corresponds to 
the conductive type between two CrO$_2$ particles, while the second to the 
non-conductive type between two Cr$_2$O$_3$ particles or along a mixed
CrO$_2$-Cr$_2$O$_3$ link, neglecting any possible leakage current through 
Cr$_2$O$_3$.

At the same time, each jump of the walker covers an amount of displacement,
measured between the geometric centers of the particles, which is tracked and
used to calculate the mean-square displacement $\langle r^{2}(t) \rangle$ as
a function of total time $t$ spent by the walker.
(We have also evaluated $\langle r^{2}(t) \rangle$ based on
the displacements between interparticle contacts, and obtained very similar
results.) In our simulations the progress 
of each random walk was tracked for $10^6$ steps through 
the system. The diffusion constant $D$ can then be calculated via the relation
$\langle r^{2}(t) \rangle \propto Dt$. It is, in turn, linearly proportional to the
dc conductivity $\sigma$ of the system via the Nernst-Einstein relation
\cite{HavlinAdvPhys87}
\begin{equation}
\label{nernst}\sigma = \left ( e^{2}/k_{B} T \right ) n D,
\end{equation}
where $n$ is the density of the charge carriers. We can then map the behavior
of the electrical conductivity of complex networks onto the diffusion property
of random walks. In the disorderly packed system of binary composites studied
here, the carrier density $n$ is determined by the concentration of conductive
(CrO$_2$) particles with fraction $p$; thus it can be approximated as
$n \sim pf$, where $f$ is the packing fraction of all the particles (CrO$_2$
and Cr$_2$O$_3$) evaluated from simulations. In our calculations for each
system size, this procedure was repeated with $20$ different randomly
selected starting particles (i.e., $20$ independent walks) in each simulated
configuration of binary spherocylinders, and the results were averaged over
a large number of system configurations generated independently.

\subsubsection{Finite size scaling}

Continuum percolation models for disordered particle systems exhibit the same
form of critical phenomena as the more-studied lattice percolation models by
virtue of universality. This critical behavior is characterized by power-law
scaling relations of geometrical, statistical, or some functional properties
in terms of the particle probability or fraction $p$ near the percolation
threshold $p_{c}$ \cite{Stauffer}, such as the scaling of conductivity
$\sigma \propto (p - p_{c})^\mu$ when $p \rightarrow p_c$ as examined here
for the metal-insulator nanocomposite. Another important quantity is
the correlation length $\xi$, scaling as $\xi \propto |p-p_{c}|^{-\nu}$,
which represents the characteristic size of the finite clusters
\cite{Stauffer,Diffuse}.

In principle, all such critical behaviors are defined in the limit of infinite
system sizes, requiring simulations to be related to the infinite system to
obtain accurate scaling relations and critical exponents. It is particularly
important for an accurate determination of $p_{c}$, because it defines the reference
point of all the scaling relationships for the other critical properties of the
system. The definition of $p_{c}$ is the particle concentration or fraction
at which the infinite system is first able to generate an infinite, percolating
cluster. However, for finite systems what is measured instead is the effective
percolation threshold as a function of the specific finite system size $L$,
i.e., $p_{c}^{\rm eff}(L)$. Thus the finite size scaling is needed to extract
the infinite-system results of $p_c = p_{c}^{\rm eff}(L \rightarrow \infty)$
and critical exponents from the finite-size simulations, for which we adopt
the method of Ref.~\cite{TorquatoExp} as outlined below.

In practice, for each system size $L$, at a given value of $p$ the probability
$P(p,L)$ of getting at least one cluster of conducting particles that spans
through the system across two sides of any direction is first evaluated over
a number of independent realizations of system configurations (in the random 
packing state prepared by the mechanical contraction and MC methods
described above). The value of $p_{c}^{\rm eff}(L)$ can then be calculated 
by determining the value of $p$ at which this spanning probability
$P(p=p_{c}^{\rm eff}, L)=1/2$. $P(p,L)$ of a finite-size system is expected 
to follow the scaling relation \cite{Stauffer}
\begin{equation}
  P(p,L) = \Phi \left [ (p - p_c)/\Delta(L) \right ],
  \label{eq:P_scaling}
\end{equation}
where $\Phi$ is the scaling function, and $\Delta$ is the percolation transition 
width which tends to decrease towards 0 as $L$ increases towards the thermodynamic 
limit, typical of phase transitions. The percolation transition 
width scales as \cite{Stauffer,TorquatoExp}
\begin{equation}
  \Delta(L) \propto L^{-1/\nu},
  \label{eq:Delta}
\end{equation}
which provides a simple way to determine the correlation-length
exponent $\nu$, leading, together with Eq.~(\ref{eq:P_scaling}), to
\begin{equation}
  p_{c}^{\rm eff}(L)-p_{c} \propto L^{-1/\nu},
  \label{eq:pc_scaling}
\end{equation}
a scaling relation that is used to obtain the accurate result of the
percolation threshold $p_{c}$.

In the calculations this is done by fitting $P(p,L)$ with a function of the 
sigmoidal form. In Ref.~\cite{TorquatoExp} for a system of spheres, the scaling
fitting function was chosen as $\{1 + {\rm erf}[(p-p_{c}^{\rm eff}(L))/\Delta(L)]\}/2$,
i.e., the cumulative distribution function (CDF) for the normal distribution.
In the currently studied system consisting of highly anisotropic spherocylinders,
the probability distribution $P(p,L)$ does not have an antisymmetric form with 
respect to $P=1/2$, as seen from our simulation results; thus, here the 
fitting function is assumed as the CDF for the skew normal (SN) distribution 
$\Phi_{\rm SN}[(p-\xi)/\omega, \alpha]$, where the location, scale, and shape
parameters of the SN distribution are denoted by $\xi$, $\omega$, and $\alpha$,
respectively. If $\alpha=0$ the CDF for the normal distribution (as used in 
Ref.~\cite{TorquatoExp}) is recovered. For each system size $L$, defining 
$p_c^{\rm eff}$ as the value when $\Phi_{\rm SN}(p=p_c^{\rm eff})=1/2$ and 
$y_0=(p_c^{\rm eff}-\xi)/\omega$, we have $\Phi_{\rm SN}[(p-\xi)/\omega, \alpha] 
= \Phi_{\rm SN}[(p-p_c^{\rm eff})/\omega+y_0, \alpha]$, with very similar value 
of $y_0$ obtained from our simulation results of different $L$. Thus when setting 
$\omega \equiv \Delta(L)$, $P(p,L)$ can be fitted to the scaling function
\begin{equation}
  P(p, L) = \Phi_{\rm SN}\left[\frac{p-p_c^{\rm eff}(L)}{\Delta(L)}, \alpha(L)
  \right ]. \label{eq:P_SN}
\end{equation}
We have applied both fitting functions, i.e., CDFs for normal and skew normal
distributions, to the simulation data, and obtained very similar results of $p_c$
(within numerical errors) after performing the finite size scaling of 
Eq.~(\ref{eq:pc_scaling}). However, the CDF $\Phi_{\rm SN}$ for the skew normal 
function is a better fit for $P(p,L)$ curves than the CDF for the normal 
distribution. Therefore, in the following only the results obtained from fitting 
to Eq.~(\ref{eq:P_SN}) are presented.

To summarize, by running a series of trials of various values of $p$ across the
transition regime near $p_{c}^{\rm eff}$ for a range of system sizes $L$, we can first
evaluate $\Delta(L)$ and $p_{c}^{\rm eff}$ through the fitting to Eq.~(\ref{eq:P_SN})
and directly measure how they scale as a function of $L$. The next step is to use
these scaling relations to identify the value of $\nu$ from Eq.~(\ref{eq:Delta})
and then, based on Eq.~(\ref{eq:pc_scaling}), extrapolate to the infinite system
to obtain the true percolation threshold $p_{c}$. 
The precise value of $p_c$ identified from Eq.~(\ref{eq:pc_scaling})
can then be used to determine other critical exponents via scaling relations,
particularly the critical exponent $\mu$ for conductivity $\sigma \propto (p-p_c)^\mu$
calculated from the random walk (de Gennes ant) method described above.

\subsection{Simulation results}

Using the methods described above, we conducted a series of simulations of
randomly packed conducting-insulating spherocylinders, for six systems with total
number of particles $N=1728$, $2744$, $4096$, $5832$, $8000$, and $10648$.
In each system, the particle number is of the form $N=m^{3}$ where $m$ is an
integer ranging from $12$ to $22$, to allow for simple initialization in a
cubic-shape simulation box with periodic boundary conditions. 
All the spherocylinders were taken to be of the same size, 
with $0.16$ units in diameter $d$ and $1.04$ units in length, corresponding to 
an aspect ratio of $6.5$, same as that of CrO$_2$/Cr$_2$O$_3$ nanocomposites. 
A typical simulation started with a cubic box of edge length $2.5m$, to 
ensure sufficient spacing between the spherocylinders for the initial 
randomization steps. The positions and orientations of the particles were 
initially thermalized with $10^4$ Hard-Particle Monte Carlo steps at a density 
$<0.01$. Then the MCM algorithm described above was applied to generate the 
final dense disordered state with the corresponding compressed system size $L$. 
For the example of $22^3 = 10648$ particles, the fully 
compressed system has an edge length of around 9 units, with $L/d=56.54$.
The list of contacting particles obtained during the steps of the MCM 
routines for overlap removal
was utilized to identify connected clusters and percolation paths of the final
dense packing state through the Hoshen-Kopelman algorithm \cite{HoshenPRB76}
with off-lattice extension \cite{Al-Futaisi03}. For a given system, at each
fraction $p$ of conducting particles, we performed a large number of
replicate simulations, $\mathcal{R}$, with different initial random number
seeds, to generate different configurations of dense random
packing through MCM for each $p$ to assure the results were independent.
For our initial hypothesis exploration, the number of replicates was chosen 
in such a way that the expected relative statistical error
$\sqrt{N}/N+\sqrt{\mathcal{R}}/\mathcal{R}<0.1$. 
We then refined our data by adding approximately the same number of runs for 
each $p$, bringing the total statistical error close to $0.05$. 
For example, for a system of $N = 22^3 = 10648$ particles, we generated $355$ 
independent configurations via MCM to evaluate the system property 
at each $p$, while for $N=12^3=1728$, we used $530$ replicates for each $p$.

\begin{figure}
\begin{center}
\includegraphics[width=0.49\textwidth]{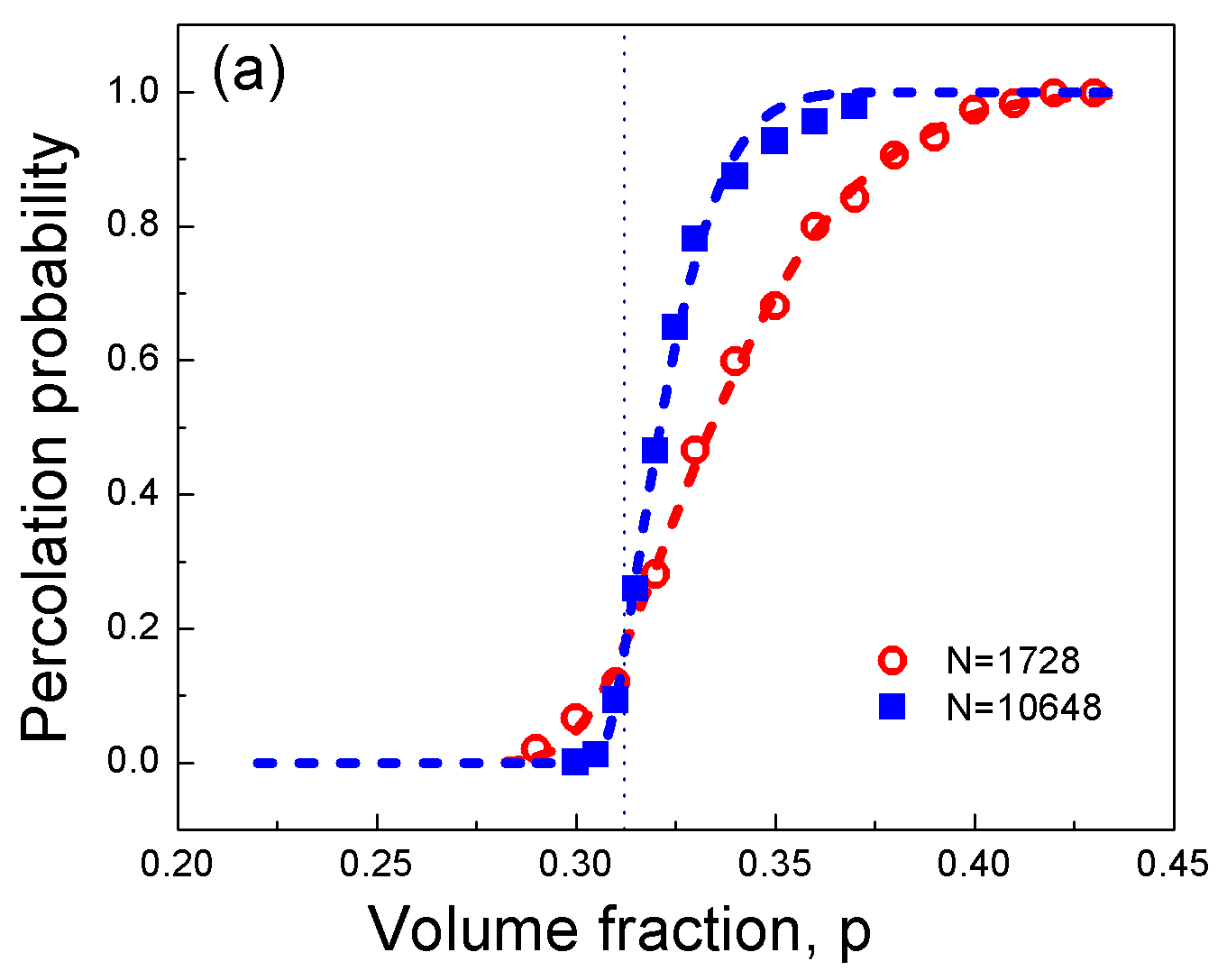}
\includegraphics[width=0.49\textwidth]{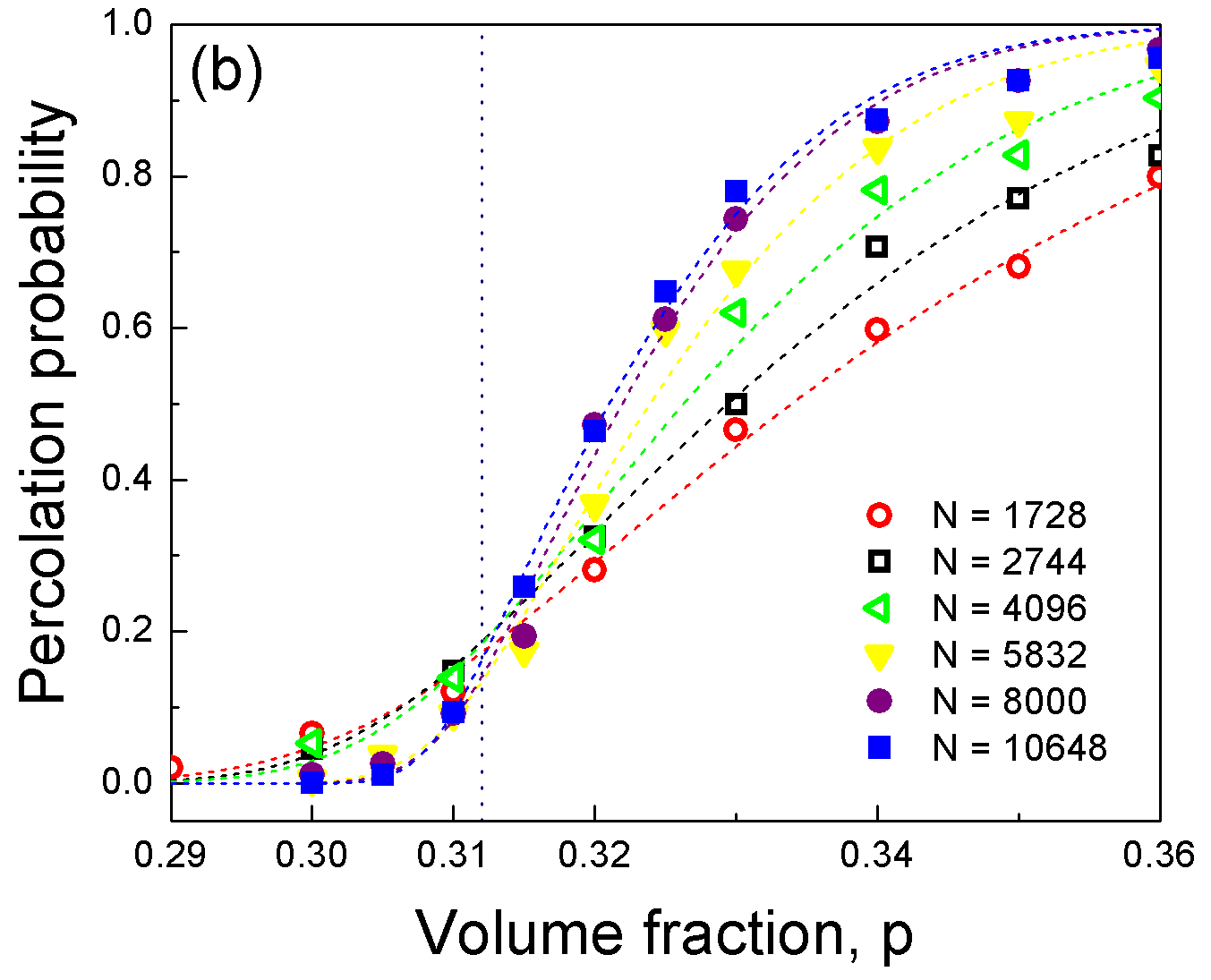}
\caption{Spanning percolation probability as a function of fraction $p$
  of the conducting spherocylinders, for (a) two sample systems with $N=1728$
  (open circles) and $N=10648$ (filled squares) particles, and 
  (b) an enlarged portion for all six systems of different $N$. 
  Each set of simulation data (shown as symbols) is fitted to Eq.~(\ref{eq:P_SN}), 
  with results shown as dashed curves. The vertical dashed line in both (a) and
  (b) indicates the location of $p_c=0.312$.}
\label{P-p}
\end{center}
\end{figure}

\begin{figure}
\begin{center}
\includegraphics[width=0.49\textwidth]{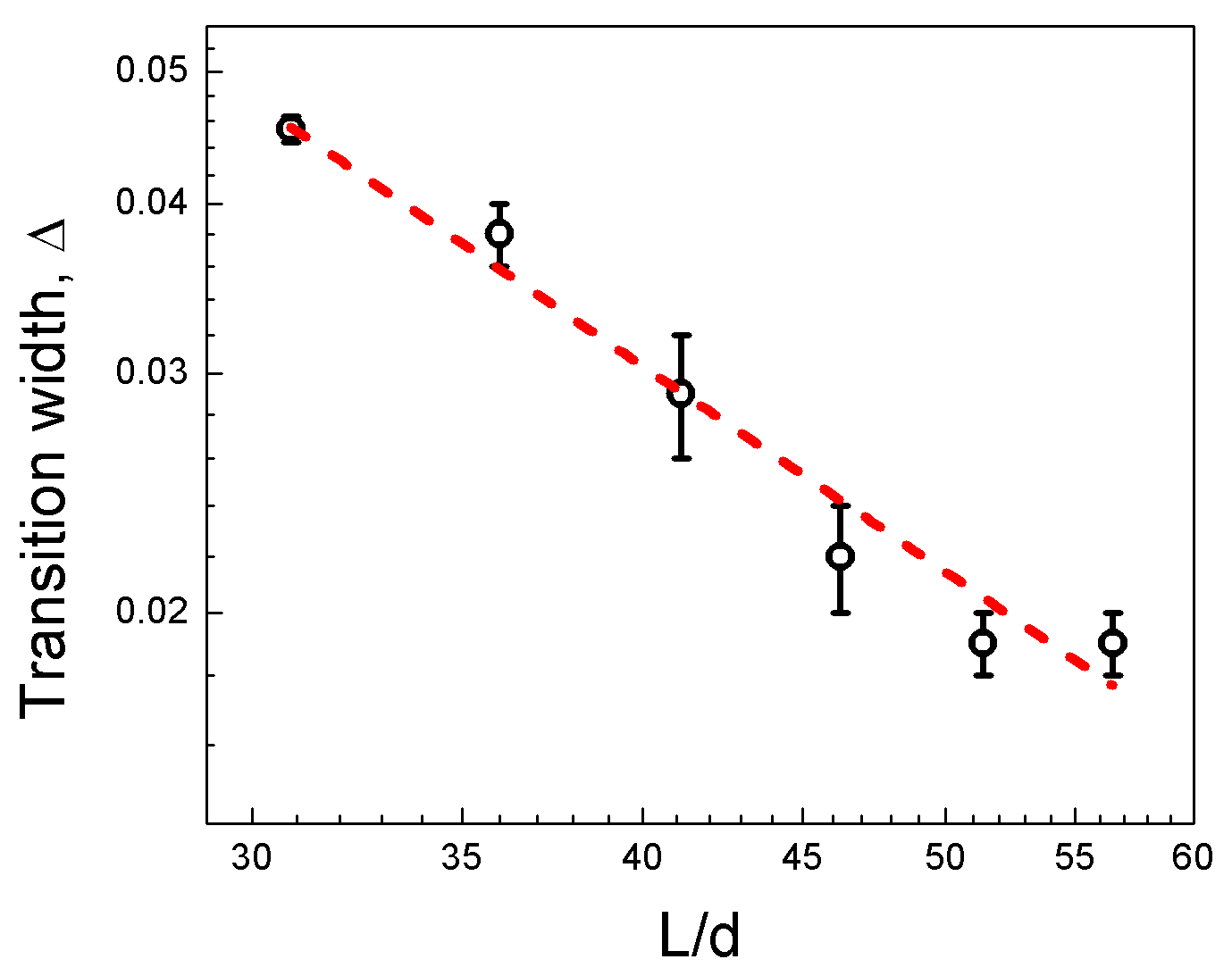}
\caption{Width $\Delta$ of the percolation transition as a function of system
  size $L$ rescaled by the particle diameter $d$. The error bars are from
  the fitting of Eq.~(\ref{eq:P_SN}) to obtain $\Delta$ at each $L$. The dashed
  line is the power-law fit to Eq.~(\ref{eq:Delta}), yielding a value of the
  critical exponent $\nu=0.646 \pm 0.005$.}
\label{width_L}
\end{center}
\end{figure}

\begin{figure}
\begin{center}
\includegraphics[width=0.49\textwidth]{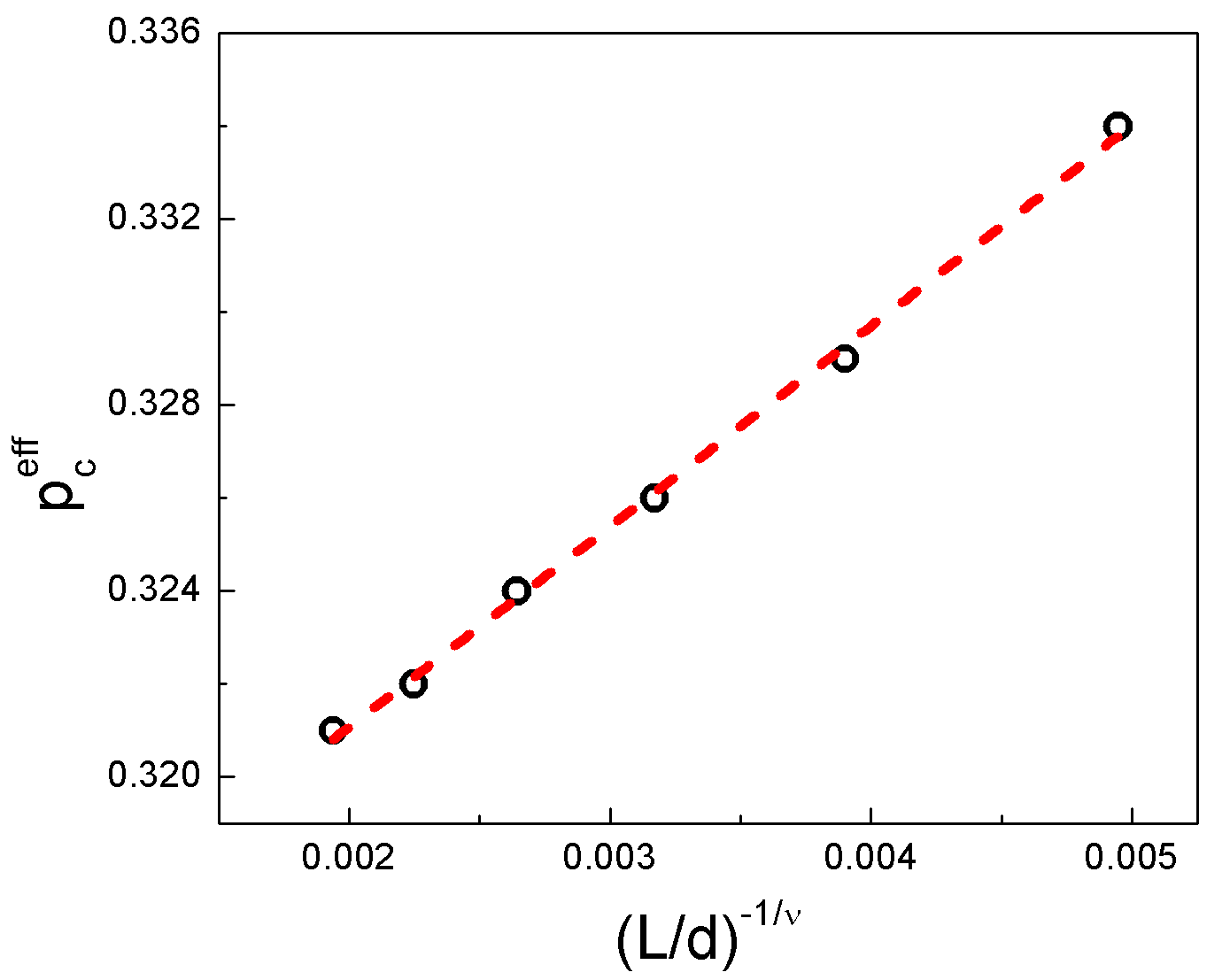}
\caption{Effective percolation threshold $p_c^{\rm eff}$ as a function of
  $(L/d)^{-1/\nu}$. The dashed line is the fit to
  Eq.~(\ref{eq:pc_scaling}), and its intersection with the $y$ axis gives the
  percolation threshold $p_{c}=0.312 \pm 0.002$ for the infinite system
  with $L \rightarrow \infty$.}
\label{pc_eff}
\end{center}
\end{figure}

Figure \ref{P-p}(a) shows the plots of the spanning percolation probability $P$
calculated over a range of $p$ near the percolation threshold,
for two different particle numbers, $N = 1728$ and $N = 10648$. A smaller $N$ 
(corresponding to a smaller system size $L$) yields a broader probability 
distribution, with a larger width of percolation transition $\Delta$, as expected. 
The quantitative values of $\Delta$ and the effective percolation threshold 
$p_c^{\rm eff}$ for various system sizes $L$ were determined through the fits 
of the calculated probability data to Eq.~(\ref{eq:P_SN}), with results 
presented in Figs.~\ref{width_L} and \ref{pc_eff}, respectively.

Following the finite-size scaling procedure described above, we first determine 
the value of the correlation-length critical exponent $\nu$ by fitting the values 
of transition width $\Delta$ to Eq.~(\ref{eq:Delta}), as shown in Fig.~\ref{width_L}.
It gives $\nu = 0.646 \pm 0.005$ which, in turn, is used to fit the values of 
$p_{c}^{\rm eff}$ plotted against $L^{-1/\nu}$ according to Eq.~(\ref{eq:pc_scaling}). 
The fitting result, presented in Fig.~\ref{pc_eff}, is used to determine the 
extrapolated value of the percolation threshold $p_{c}=0.312 \pm 0.002$ in the 
limit of infinitely large system of randomly packed spherocylinders. 
This value of $p_c$ agrees well with the experimental finding of $p_{c}=0.305 \pm 0.026$ 
reported in Sec.~\ref{sec:experiment_results} for CrO$_2$/Cr$_2$O$_3$ nanocomposites.
We note that the value of correlation-length exponent $\nu$ ($=0.646$)
obtained here is smaller than the standard value of $0.876$ in 3D, which could be
attributed to the relatively large errors of width $\Delta$ evaluated in the fitting
(see Fig.~\ref{width_L}) and limited number of particles used in our simulations.

An alternative method to identify $p_c$ can be used by noting that
given a certain geometry of the constituent particles and a dimensionality of the
system, the critical spanning probability at $p=p_c$ is a universal quantity
\cite{Thr1}. Thus, for systems of different finite sizes the corresponding spanning
probabilities $P(p,L)$ are expected to cross at $p=p_c$. This is consistent with our
simulation results given in Fig.~\ref{P-p}(b), which shows a narrow range of such 
crossing points located from $p=0.307$ to $0.317$ due to numerical variations, 
well agreeing with the result of $p_{c}=0.312$ obtained above from finite size
scaling.

\begin{figure}
\begin{center}
\includegraphics[width=0.49\textwidth]{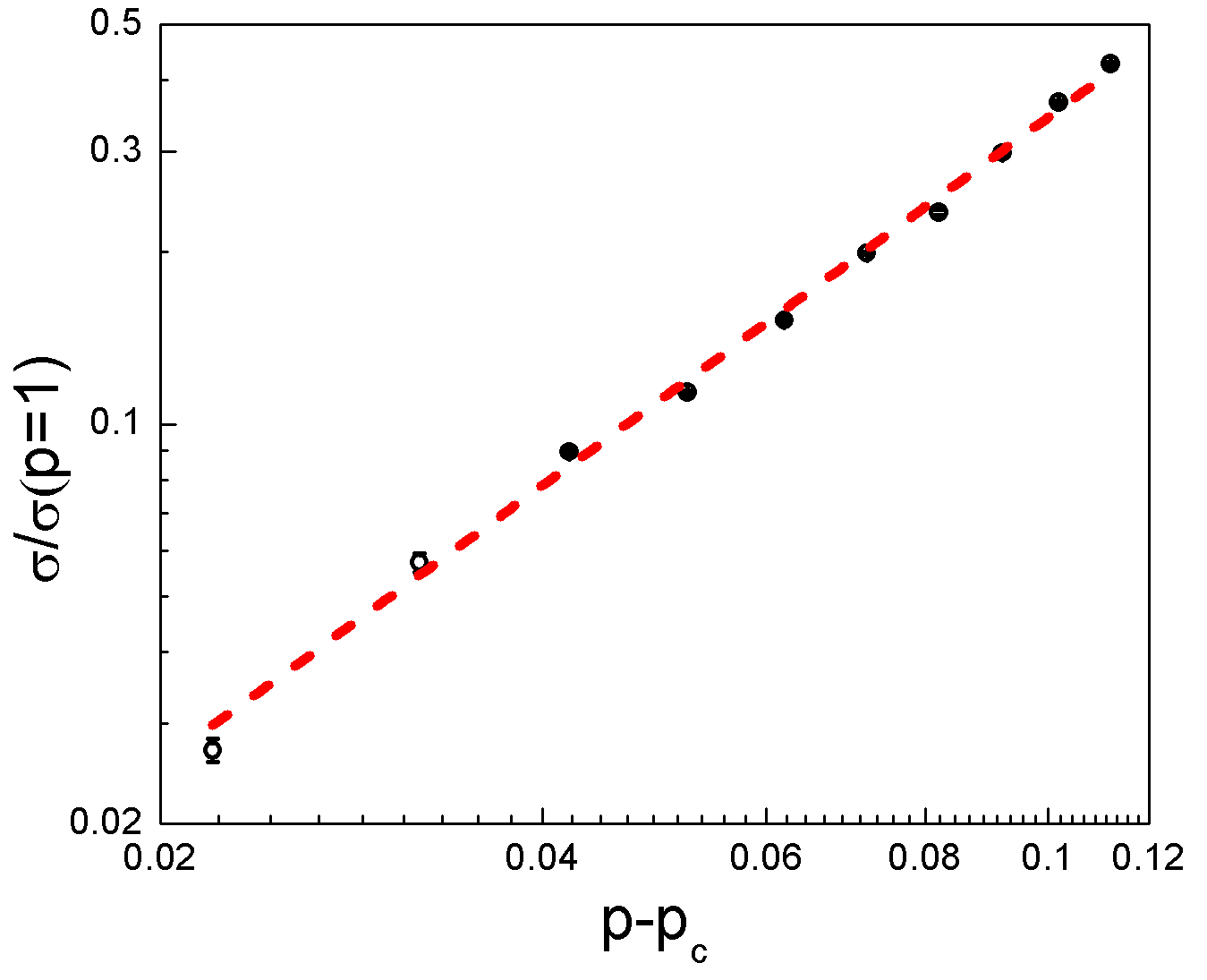}
\caption{Rescaled conductivity, $\sigma(p)/\sigma(p=1)$, as a 
  function of $p-p_c$ for systems of $N=10648$ particles. The fitting to the 
  scaling relation $\sigma \propto (p-p_c)^\mu$ (shown as the dashed line)
  gives the conductivity critical exponent $\mu=1.62 \pm 0.04$.}
\label{Conductivity}
\end{center}
\end{figure}

Finally, we ran another series of simulations and applied the random 
walk method of de Gennes ant described above to calculate the system conductivity 
$\sigma$ for a narrow range of $p$ above $p_c$. A large system size with 
$N=22^3=10648$ particles was used, with $350$ independent configurations generated
via MCM for each $p$. In each configuration $20$ independent random walks of 
``blind ants'' (with randomly chosen starting points) were conducted, and the 
outcomes were averaged over $350\times20=7000$ trials at each $p$ to obtain the 
mean-square displacement $\langle r^{2}(t) \rangle$. The asymptotic regime of 
$t=5\times 10^5 - 10^6$ was used to calculate the system conductivity $\sigma$.
The corresponding results of $\sigma$ are presented in Fig.~\ref{Conductivity}, 
where the conductivity data has been rescaled with respect to $\sigma(p=1)$, 
the conductivity of the system consisting of purely conducting particles that 
was evaluated at $N=10648$. Using this data to fit into the scaling relation
$\sigma \propto (p-p_c)^\mu$, we determined the critical exponent $\mu$ of 
the system conductivity, yielding $\mu=1.62 \pm 0.04$.

This calculated value of the conductivity critical exponent is within the range of 
previous findings of $1.3 \leq t \leq 4.0$ in disordered composites 
\cite{BauhoferCST09,Vionnet-MenotPRB05,BalbergPRL17}, but much smaller than the 
value of $t=2.52$ measured for this system experimentally. This discrepancy may be 
understood in light of analysis of Ref.~\cite{BalbergPRL17}, given that in the
nanocomposite system studied here the electrical transport between any two adjacent 
spherocylinders is mainly governed by their interfacial resistance (due to the 
presence of surface oxide layers). Thus, the scaling of the system conductivity 
should be written as \cite{BalbergPRL17}
\begin{equation}
    \sigma \propto (p_j - p_{jc})^\mu \propto (p^2 - p_c^2)^\mu 
    = (p-p_c)^\mu (p+p_c)^\mu,
    \label{eq:sigma_j}
\end{equation}
where $p_j$ is the occupation probability of the conducting interjunction (proportional
to the average contact number between conducting particles), and $p_j \propto p^2$ with 
$p$ the fraction or concentration of conducting particles, as confirmed in our numerical
simulations. When $p$ is close to $p_c$ (as in our computation 
of $\sigma$ shown in Fig.~\ref{Conductivity}), from Eq.~(\ref{eq:sigma_j}) the 
conductivity scaling behavior is dominated by $\sigma \propto (p-p_c)^\mu$, 
recovering the standard scaling relation for $\sigma$, with the value of 
$\mu=1.62 \pm 0.04$ identified above. On the other hand, when $p$ is far enough from the 
percolation threshold $p_c$, as occurred in our experimental measurement and data analysis 
(see Fig.~\ref{R-p}), the range of $\mu < t \leq 2\mu$ for the conductivity exponent 
$t$, as measured for $\sigma \propto (p-p_c)^t$, is expected \cite{BalbergPRL17}, 
leading to a lower bound of $\mu \geq \mu_l = t/2 = 1.26$ for the experimental result 
$t = 2.52 \pm 0.03$ given in Sec.~\ref{sec:experiment_results}.
Our computed result of $\mu$ is then consistent with the experimentally measured value 
for the critical exponent, suggesting that our model correctly captures the essential 
electrical property of this percolating network of disordered spherocylinders.

\section{Discussion and Conclusions}

We have investigated a binary composite system of randomly packed spherocylinders, 
to examine the relation between structural (percolation) and transport properties 
of the system. The composite consists of metal (CrO$_2$) - insulator (Cr$_2$O$_3$) 
nanoparticles that are of identical rodlike particle geometry but distinct 
functionality. Our experimental ($p_{c}=0.305 \pm 0.026$, $t=2.52 \pm 0.03$ with
$\mu_l=1.26$) and computational 
($p_{c}=0.312 \pm 0.002$, $\mu=1.62 \pm 0.04$) results for both the 
percolation threshold and conductivity critical exponent are in good agreement. 
The small observed variations can be partially attributed to a different degree of 
disorder in experimental and theoretical arrangement of the spherocylinders. 
While in our simulations we have used a completely disordered system, 
the samples fabricated for the experimental measurements have a substantial 
degree of local nematic order which, in turn, may affect the threshold 
and the critical exponent values. The effects of polydispersity 
and particle irregularity, which are unavoidable in the experimental setup 
(see Fig.~\ref{Lemons}) but neglected in the simulations, may also account 
for the discrepancy between the experimental and computational results.
In addition, the experimental results near the classical percolation 
threshold $p_c$ may have been affected by the presence of tunneling percolation 
at lower values of $p$.

We note that our large-scale simulation result for the site percolation threshold
of spherocylinders with an aspect ratio of $6.5$, i.e., $p_c=0.312 \pm 0.002$, is 
very close, within numerical errors, to the threshold found in Ref.~\cite{Z-T} for
jammed disordered spheres [$p_c=0.3116(3)$] which, in turn, is almost identical to
the threshold of the simple cubic lattice ($p_c=0.311608$ \cite{Thr1,Thr3}).
These three systems are geometrically quite distinct, with very
different packing factors and different distribution of coordination numbers,
although their average coordination numbers $\bar{Z}$ are close. The coordination
number for the simple cubic lattice is exactly 6, the disordered jammed sphere
packing has coordination numbers ranging from 4 to 12, with an average of 6
\cite{Z-T}, while for the random close-packing of spherocylinders with a $6.5$
aspect ratio studied here, a broader distribution of coordination numbers is
obtained, ranging from 0 to 16 with an average of $\bar{Z} = 5.83 \pm 0.07$ 
(averaged over $200$ simulations with $10648$ particles each).

This value of $\bar{Z}$ is much lower than the isostatic value
of $\bar{Z}=2d_f=10$ for spherocylindrical particles with $d_f=5$ degrees of 
freedom per particle \cite{DonevPRE07} at sufficiently large aspect ratios, 
indicating that the randomly packed systems examined here are not strictly jammed. 
It is likely due to the use of MCM, which has been known to produce states of
lower $\bar{Z}$ for long rods with high aspect ratio \cite{WouterseGM09} and 
the use of insufficiently small contact tolerance in our numerical simulations 
as compared to that generating random jammed packings of spheres 
\cite{Z-T,AtkinsonPRE13}. This results in a disordered packing state of 
spherocylinders that are not fully jammed, similar to the nanocomposite system 
studied here experimentally for which the strict jamming with higher $\bar{Z}$ 
is usually not accessible.

Although systems with the same average coordination number $\bar{Z}$ are likely 
to have similar values of $p_c$, such a close agreement is unexpected, given 
the different distribution of coordination numbers and noting that some other 
systems with the same average coordination number of 6 do not have such close 
values of $p_c$ (see Table 1 of Ref.~\cite{Z-T}).
This result implies that not only the values of the threshold may be unaffected 
by the exact details of particle ordering, as in the case of ordered lattice 
vs disordered packing of spheres with the same average coordination number, 
as has been pointed out in Ref.~\cite{Z-T}, but it may also be insensitive 
to the details of the particle geometric shape.
While this may still be fortuitous, the coincidence seems less likely,
given the independent research on three different systems, and with the 
new result presented here for the dense disordered state of spherocylinders 
that are geometrically distinct from spheres. If this result also holds for 
other systems, it may indicate a possible \textit{universality} of the 
percolation threshold based on a more profound underlying mechanism which 
is presently unknown and needs further investigation.

These results are relevant for future device applications of functional nanoparticle
composites, for which the ability to control the percolation threshold is critically
important. They can also be of interest for a broader range of packing and percolation
systems, such as drug release and the design of drug tablets \cite{DrugSimulations},
for which the percolation of soluble drug nanoparticles in the packed soluble/insoluble
composite plays an important role \cite{Z-T}.

\begin{acknowledgments}
Z.-F.H. acknowledges support from the National Science Foundation under Grant 
No.~DMR-1609625. The authors are grateful to Isaac Balberg, Robert Ziff, and
Salvatore Torquato for illuminating discussions and helpful suggestions. 
\end{acknowledgments}


\begin{thebibliography}{99}

\bibitem {Stauffer} D. Stauffer and A. Aharony, {Introduction to
Percolation Theory} (Taylor \& Francis, London, 1994).

\bibitem {epidemic} R. Parshani, S. Carmi, and S. Havlin, Epidemic
threshold for the susceptible-infectious-susceptible model on random
networks, Phys. Rev. Lett. \textbf{104}, 258701 (2010).

\bibitem {thermal} P. Bonnet, D. Sireude, B. Garnier, and O. Chauveta, 
Thermal properties and percolation in carbon nanotube-polymer composites, 
Appl. Phys. Lett. \textbf{91}, 201910 (2007).

\bibitem {Diffuse} A. Bunde and W. Dieterich, Percolation in
Composites, J. Electroceram. \textbf{5}, 81 (2000).

\bibitem {electrical} Z. Ball, H. M. Phillips, D. L. Callahan, and R.
Sauerbrey, Percolative metal-insulator transition in excimer laser irradiated
polyimide, Phys. Rev. Lett. \textbf{73}, 2099 (1994).

\bibitem {magnetic} H. J. Elmers, J. Hauschild, H. H\"{o}che, U. Gradmann, H.
Bethge, D. Heuer, and U. K\"{o}hler, Submonolayer Magnetism of Fe(110) on
W(110): Finite Width Scaling of Stripes and Percolation between Islands, Phys.
Rev. Lett. \textbf{73}, 898 (1994).

\bibitem {spinHall} I. A. Gruzberg, A. W. W. Ludwig, and N. Read,
Exact Exponents for the Spin Quantum Hall Transition, Phys. Rev. Lett. 
\textbf{82}, 4524 (1999).

\bibitem {Pharma1} R. A. Siegel, J. Kost, and R. Langer, Mechanistic
studies of macromolecular drug release from macroporous polymers, J.
Control Release \textbf{8}, 223 (1989).

\bibitem {Pharma2} I. Caraballo, M. Fernandez-Arevalo, M.A. Holgado, and A.M.
Rabasco, Percolation theory: application to the study of the release behaviour
from inert matrix systems, Int. J. Pharm. \textbf{96}, 175 (1993).

\bibitem{BauhoferCST09} W. Bauhofer and J. Z. Kovacs, A review and analysis of electrical
percolation in carbon nanotube polymer composites, Compos. Sci. Techno. \textbf{69},
1486 (2009).

\bibitem{Vionnet-MenotPRB05} S. Vionnet-Menot, C. Grimaldi, T. Maeder,
S. Str\"{a}ssler, and P. Ryser, Tunneling-percolation origin of nonuniversality: 
Theory and experiments, Phys. Rev. B \textbf{71}, 064201 (2005).

\bibitem{BalbergPRL17} I. Balberg, Unified Model for Pseudononuniversal Behavior of 
the Electrical Conductivity in Percolation Systems, Phys. Rev. Lett. \textbf{119}, 
080601 (2017).

\bibitem {Jamming} S. Torquato and F. H. Stillinger, Jammed hard-particle
packings: From Kepler to Bernal and beyond, Rev. Mod. Phys. \textbf{82}, 2633 (2010).

\bibitem {double} X. Liu, R. P. Panguluri, Z.-F. Huang, and B. Nadgorny, 
Double Percolation Transition in Superconductor-Ferromagnet Nanocomposites, 
Phys. Rev. Lett. \textbf {104}, 035701 (2010).

\bibitem {dJ-B} M. J. M. de Jong and C. W. J. Beenakker, Andreev Reflection in 
Ferromagnet-Superconductor Junctions, Phys. Rev. Lett. \textbf {74}, 1657 (1995).

\bibitem {Andreev} R. J. Soulen Jr., J. M. Byers, M. S. Osofsky, B. Nadgorny,
T. Ambrose, S. F. Cheng, P. R. Broussard, C. T. Tanaka, J. Nowak, J. S. Moodera, 
A. Barry, and J. M. D. Coey, Measuring the spin polarization of a metal with a 
superconducting point contact, Science \textbf{282}, 85 (1998).

\bibitem{Thr1} C. D. Lorenz and R. M. Ziff, Universality of the excess
number of clusters and the crossing probability function in three-dimensional
percolation, J. Phys. A: Math. Gen. \textbf{31}, 8147 (1998).

\bibitem {Thr2} H. G. Ballesteros, L. A. Fernandez, V. Martin-Mayor, A. Munoz
Sudupe, G. Parisi, and J. J. Ruiz-Lorenzo, Scaling corrections: site
percolation and Ising model in three dimensions, J. Phys. A: Math. Gen.
\textbf {32}, 1 (1999).

\bibitem {Thr3} J. Wang, Z. Zhou, W. Zhang, T. M. Garoni, and
Y. Deng, Bond and site percolation in three dimensions, Phys. Rev. E 
\textbf{87}, 052107 (2013).

\bibitem {Thr4} K. Malarz, Simple cubic random-site percolation
thresholds for neighborhoods containing fourth-nearest neighbors,
Phys. Rev. E \textbf{91}, 043301 (2015).

\bibitem {Thr5} J. Tran, T. Yoo, S. Stahlheber, and A. Small,
Percolation thresholds on three-dimensional lattices with three nearest
neighbors, J. Stat. Mech.: Th. Exp. \textbf{2013}, 05014 (2013).

\bibitem {Z-T} R. M. Ziff and S. Torquato, Percolation of disordered
jammed sphere packings, J. Phys. A: Math. Theor. \textbf{50}, 085001 (2017).

\bibitem {Annealing} J. Wang, P. Che, J. Feng, M. Lu, J. Liu, and J. Meng, 
A large low-field tunneling magnetoresistance of CrO$_{2}$/(CrO$_{2}$/Cr$_{2}$O$_{3}$) 
powder compact with two coercivities, J. Appl. Phys. \textbf {97}, 073907 (2005).

\bibitem {Fourprobe} I. Miccoli, F. Edler, H. Pfnur, and C. Tegenkamp, The
100th anniversary of the four-point probe technique: the role of probe
geometries in isotropic and anisotropic systems, J. Phys.: Condens. Matter 
\textbf{27}, 223201 (2015).

\bibitem {Stairs_Bal} I. Balberg, D. Azulay, Y. Goldstein, J. Jedrzejewski, 
G. Ravid, and E. Savir, Eur. Phys. J. B \textbf{86}, 428 (2013).

\bibitem {Stairs} R. Mukherjee, Z.-F. Huang, and B. Nadgorny, Multiple percolation 
tunneling staircase in metal-semiconductor nanoparticle composites, 
Appl. Phys. Lett. \textbf{105}, 173104 (2014).

\bibitem{ChenCPL03} Y.-J. Chen, X.-Y. Zhang, T.-Y. Cai, and Z.-Y. Li,
Hopping and non-universal conductivity in half-metallic CrO$_2$ composites,
Chin. Phys. Lett. \textbf{20}, 721 (2003).

\bibitem {MD1997} M. Kubo, Y. Omi, R. Miura, A. Stirling,
A. Miyamoto, M. Kawasaki, M. Yoshimoto, and H. Koinuma,
Atomic control of layer-by-layer epitaxial growth on SrTiO$_{3}$ (001):
Molecular-dynamics simulations, Phys. Rev. B \textbf{56}, 13535 (1997)

\bibitem {MC2006} G. Russo and P. Smereka, Computation of strained epitaxial growth 
in three dimensions by kinetic Monte Carlo, J. Comput. Phys. \textbf{214},809 (2006).

\bibitem {Metro} N. Metropolis, A. W. Rosenbluth, M. N. Rosenbluth, A. H. Teller, 
and E. Teller, Equation of State Calculations by Fast Computing Machines, 
J. Chem. Phys. \textbf{21}, 1087 (1953).

\bibitem {MCM} S. R. Williams and A. P. Philipse, Random packings of
spheres and spherocylinders simulated by mechanical contraction, 
Phys. Rev. E \textbf{67}, 051301 (2003).

\bibitem {OtherMCM} A. V. Kyrylyuk, M. A. van de Haar, L. Rossi,
A. Wouterse, and A. P. Philipse, Isochoric ideality in jammed
random packings of non-spherical granular matter, Soft Matter
\textbf{7}, 1671 (2011).

\bibitem {HOOMD1} J. A. Anderson, C. D. Lorenz, and A. Travesset,
General purpose molecular dynamics simulations fully implemented on
graphics processing units, J. Comput. Phys. \textbf{227}, 5342 (2008).

\bibitem {HOOMD2} J. Glaser, T. D. Nguyen, J. A. Anderson, P. Liu, F. Spiga, 
J. A. Millan, D. C. Morse, and S. C. Glotzer, Strong scaling of
general-purpose molecular dynamics simulations on GPUs, Comput. Phys.
Commun. \textbf{192}, 97 (2015).

\bibitem {HOOMD_HPMC} J. A. Anderson, M. E. Irrgang, and S. C. Glotzer,
Scalable Metropolis Monte Carlo for simulation of hard shapes,
Comput. Phys. Commun. \textbf{204}, 21 (2016).

\bibitem{MCM_github} https://github.com/brendonwaters/Mechanical-Contraction-Method

\bibitem {ovito} A. Stukowski, Visualization and analysis of atomistic
simulation data with OVITO - the Open Visualization Tool, Modelling Simul.
Mater. Sci. Eng. \textbf{18}, 015012 (2010).

\bibitem{deGennes} P. G. de Gennes, La percolation: un concept unificateur
(Percolation a unifying concept), La Recherche \textbf{7}, 919 (1976).

\bibitem{deGennes_termite} P. G. de Gennes, Percolation: Quelques Systemes
  Nouveaux, J. Phys. (Paris) Colloq. \textbf{41}, C3-17 (1980).

\bibitem{HavlinAdvPhys87} S. Havlin and D. Ben-Avraham, Diffusion in disordered
  media, Adv. Phys. \textbf{36}, 695 (1987).

\bibitem {Diffuse2} A. Bunde, A. Coniglio, D. C. Hong, and H. E. Stanley,
Transport in a two-component randomly composite material: scaling theory 
and computer simulations of termite diffusion near the superconducting limit,
J. Phys. A: Math. Gen. \textbf{18}, L137 (1985).

\bibitem {Diffuse3} A. Bunde, W. Dieterich, and E. Roman, Dispersed Ionic
Conductors and Percolation Theory, Phys. Rev. Lett. \textbf{55}, 5 (1985).

\bibitem {Diffuse4} H. E. Roman, A. Bunde, and W. Dieterich, Conductivity of
dispersed ionic conductors: percolation model with two critical points,
Phys. Rev. B \textbf{34}, 3439 (1986).

\bibitem {MajidPRB84} I. Majid, D. Ben-Avraham, S. Havlin, and H. E. Stanley,
Exact-enumeration approach to random walks on percolation clusters in two dimensions,
Phys. Rev. B \textbf{30}, 1626 (1984).

\bibitem {TorquatoExp} M. D. Rintoul and S. Torquato, Precise determination
of the critical threshold and exponents in a three-dimensional continuum
percolation model, J. Phys. A: Math. Gen. \textbf{30}, L585 (1997).

\bibitem{HoshenPRB76} J. Hoshen and R. Kopelman, Percolation and cluster distribution
  I. Cluster multiple labeling technique and critical concentration algorithm,
  Phys. Rev. B \textbf{14}, 3438 (1976).

\bibitem{Al-Futaisi03} A. Al-Futaisi and T. W. Patzek, Extension of Hoshen-Kopelman
  algorithm to non-lattice environments, Physica A \textbf{321}, 665 (2003).

\bibitem{WouterseGM09} A. Wouterse, S. Luding, and A. P. Philipse, 
On contact numbers in random rod packings, Granul. Matter \textbf{11}, 169 (2009).

\bibitem {DonevPRE07} A. Donev, R. Connelly, F. H. Stillinger, and S. Torquato,
Underconstrained jammed packings of nonspherical hard particles: Ellipses and ellipsoids,
Phys. Rev. E \textbf{75}, 051304 (2007).

\bibitem{AtkinsonPRE13} S. Atkinson, F. H. Stillinger, and S. Torquato,
Detailed characterization of rattlers in exactly isostatic, strictly jammed sphere packings,
Phys. Rev. E \textbf{88}, 062208 (2013).

\bibitem {DrugSimulations} W. Jiang, S. Kim, X. Zhang, R. A. Lionberger, 
B. M. Davit, D. P. Conner, and L. X. Yu, The role of predictive biopharmaceutical 
modeling and simulation in drug development and regulatory evaluation, 
Int. J. Pharmaceutics \textbf {418}, 151 (2011).

\end{thebibliography}
\end{document}